%% file: Paper.tex
\pdfoutput=1
\documentclass[letterpaper,twocolumn,10pt]{article}
\usepackage{usenix2019_v3}

\input{header}

\hypersetup{
  pdftitle={Heddle: Learning Structural Templates for Parallelism Planning on Heterogeneous GPU Clusters},
  pdfauthor={Taeyoon Kim, Yonguk Song, Seoyeong Choy, Hexiao Duan, Dong Li, Seo Jin Park, Myeongjae Jeon},
}

\newcommand{\oursname}{Heddle}
\newcommand{\ours}{\textsf{\oursname}}

\begin{document}

\date{}

\title{\Large \bf Heddle: Learning Structural Templates for Parallelism \\ Planning on Heterogeneous GPU Clusters}

\author{
{\rm Taeyoon Kim$^\dagger$, Yonguk Song$^\ddagger$, Seoyeong Choy$^\ddagger$, Hexiao Duan$^\S$}\\
{\rm Dong Li$^\S$, Seo Jin Park$^\P$, Myeongjae Jeon$^\ddagger$}\\[6pt]
$^\dagger$UNIST \quad $^\ddagger$POSTECH \quad $^\S$UC Merced \quad $^\P$USC
} %

\maketitle

\newcommand{\PP}[1]{\vspace{4pt}\noindent{\bf #1.}}

\input{00_abstract}

\input{01_introduction}
\input{02_background}

\input{03_analysis}

\input{04_ideas}

\input{05_design}

\input{07_evaluation}
\input{08_discussion}

\input{09_related}

\input{10_conclusion}

{\footnotesize
\bibliographystyle{plain}
\bibliography{references}
}

\end{document}

%% file: header.tex
\usepackage{booktabs}
\usepackage{amsmath}
\usepackage{amssymb}
\usepackage{graphicx}
\usepackage{caption}
\usepackage{algorithm}
\usepackage{algpseudocode}
\usepackage{tabularx}
\usepackage{enumitem}
\usepackage{listings}
\usepackage{pifont}
\usepackage{multirow}
\usepackage{wrapfig}
\usepackage{tikz}
\usepackage{subcaption}
\usepackage{siunitx}
\usepackage{mathtools}
\usepackage{xspace}
\usepackage{array}
\usepackage{float}
\usepackage{xcolor}

\usepackage[most]{tcolorbox}

\newcommand{\cmark}{\ding{51}}%
\newcommand{\xmark}{\ding{55}}%
\newcommand{\tmark}{$\triangle$}%
\newenvironment{myitemize}%
  {\begin{itemize}
	[leftmargin=0cm,
		itemindent=.3cm,
		labelwidth=\itemindent,
		labelsep=0pt,
		parsep=3pt,
		topsep=2pt,
		itemsep=1pt,
		align=left]
  }%
  {\end{itemize}}

  {\begin{enumerate}
	[leftmargin=0cm,itemindent=.5cm,labelwidth=\itemindent,
		labelsep=0pt,
		parsep=1pt,
		topsep=1pt,
		itemsep=3pt,
		align=left]
  }%
  {\end{enumerate}}

\newenvironment{myitemize2}%
  {\begin{itemize}
    [leftmargin=.3cm,
        itemindent=0pt,
        labelwidth=.15cm,
        labelsep=.15cm,
        parsep=0pt,
        topsep=2pt,
        itemsep=1pt,
        align=left]
  }%
  {\end{itemize}}

\definecolor{perldoc@keyword}{HTML}{8B008B}   %
\definecolor{perldoc@comment}{HTML}{228B22}   %
\definecolor{perldoc@string} {HTML}{CD5555}   %
\definecolor{perldoc@number} {HTML}{B452CD}   %
\definecolor{perldoc@builtin}{HTML}{658b00}   %
\definecolor{perldoc@func}   {HTML}{008b45}   %

\lstdefinestyle{code_snippet}{
basicstyle = \ttfamily\small,
keywordstyle = \color{perldoc@keyword}\bfseries,
keywordstyle = [2]\color{perldoc@func},
keywordstyle = [3]\color{perldoc@builtin},
commentstyle = \color{perldoc@comment}\normalfont,
stringstyle  = \color{perldoc@string},
backgroundcolor = \color{white},
frame        = none,
xleftmargin  = 0em,
showstringspaces = false,
breaklines   = false,
keywords     = {for, in, if, else, return, while, def, class},
keywords     = [2]{PartitionLayers, Evaluate},
keywords     = [3]{range, len, print, type},
morecomment  = [l]{\#},
morestring   = [b]",
morestring   = [b]',
}

\newcounter{takeaway}

\newtcolorbox{takeawaybox}[2][]{
  colback=white,
  colframe=white,
  boxrule=0pt,
  arc=0pt,
  left=0pt, right=0pt, top=0pt, bottom=0pt,
  boxsep=0pt,
  before={\vspace{6pt}\noindent\refstepcounter{takeaway}%
    \ifx\\#1\\\else\label{#1}\fi},
  after={\vspace{6pt}},
  before upper={%
    \textcolor{blue}{\textbf{Takeaway \##2:}}\ \itshape
  },
}

\newcommand{\eatspace}[1]{}

\makeatletter
\newcommand{\thickhline}{%
    \noalign {\ifnum 0=`}\fi \hrule height 1pt
    \futurelet \reserved@a \@xhline
}
\newcolumntype{"}{@{\hskip\tabcolsep\vrule width 1pt\hskip\tabcolsep}}
\makeatother

\newcommand{\allnotes}[1]{}

%% file: 00_abstract.tex
\begin{abstract}
Training large machine learning models on shared GPU infrastructures faces
two challenges: (1) GPU availability shifts dynamically with varying
resource demands from tenants, and (2) hardware heterogeneity accumulates
as datacenters continuously adopt new GPU generations. Due to the vast
search space induced by heterogeneous GPU types and node sizes, training
planners must prune it aggressively to remain tractable, yet must also
derive high-throughput plans promptly as cluster configurations change.
\ours{} achieves this goal through a learning-based planner that reduces the
full planning problem to a search over pipeline structures.
\ours{} encapsulates planning decisions in a structural template and learns
to construct plans from templates over diverse cluster configurations offline.
This design is effective because structural decisions constitute the
performance-critical core of a parallelism plan, while the rest 
follows by rule or from a small priced candidate set once the plan structure is fixed.
Evaluation shows that \ours{} matches or exceeds %
the best plan found by five existing
planners across clusters with varying GPU types and node sizes for three
models of different sizes by up to 84.5\% in throughput on dense models and 4.6$\times$ on MoE models.

\end{abstract}

%% file: 01_introduction.tex
\section{Introduction}  %
\label{sec:Introduction}

Large-scale ML training draws its GPUs from public clouds~\cite{megascale,
bamboo, parcae} or shared institutional clusters~\cite{philly, MLaaS,
gpunion2025}. As each new accelerator generation joins the old ones, these
infrastructures accumulate GPU diversity: Google Cloud offers 13 NVIDIA GPU
types across seven microarchitecture generations~\cite{googlecloud_gpus},
and Alibaba's production AI cluster holds 12 GPU models from multiple
vendors across 37K servers~\cite{ASI}.
Because tenants share these GPUs and tenant demand is hard to predict, the
number of available GPUs of each type also shifts while a job is
running~\cite{gandiva-fair, gavel, MLaaS, tenplex}. A training system on such
a pool therefore needs a \emph{planner} that derives a high-throughput
parallelism plan for whatever GPUs are available now, fast enough that the
job does not stall.

\begin{table*}[t]
   \centering
   \caption{Capabilities of heterogeneous 3D
   parallelism planners.
   \cmark{} fully supported,
   \tmark{} supported with restrictions,
   \xmark{} not supported.
   \autoref{sec:motivation} analyzes each
   planner in detail.
}
   \label{tab:planner-comparison}
   \small
   \begin{tabular}{@{}lcccccc@{}}
       \toprule
       & \textbf{AMP}~\cite{amp}
       & \textbf{Sailor}~\cite{Sailor}
       & \textbf{Metis}~\cite{Metis}
       & \textbf{HexiScale}~\cite{HexiScale}
       & \textbf{Espresso}~\cite{Espresso}
       & \textbf{\oursname} \\
       \midrule
       Adapt TP degree per stage &
       \xmark & \cmark & \cmark & \cmark & \xmark & \cmark \\
       Plan expert parallelism (MoE) &
       \xmark & \xmark & \xmark & \xmark & \xmark & \cmark \\
       Utilize partial GPUs &
       \tmark & \cmark & \cmark & \cmark & \cmark & \cmark \\
       Joint device-TP search &
       \tmark & \cmark & \cmark & \tmark & \xmark & \cmark \\
       Non-uniform node sizes &
       \xmark & \xmark & \cmark & \cmark & \xmark & \cmark \\
       Unequal layer partition &
       \tmark & \xmark & \cmark & \cmark & \cmark & \cmark \\
       Scalable planning time &
       \cmark & \tmark & \xmark & \cmark & \cmark & \cmark \\
       Profiled cost estimation &
       \tmark & \cmark & \tmark & \xmark & \tmark & \cmark \\
       \bottomrule
   \end{tabular}
\end{table*}

\begin{figure}[t]
    \centering
    \includegraphics[width=\linewidth]{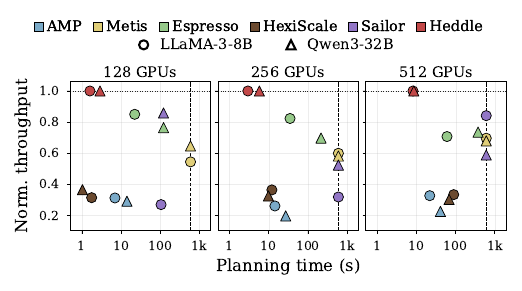}
    \caption{Training throughput achieved by competing planners versus planning time
    on a heterogeneous cluster (A100-40 + A6000 + H100 in a 1:1:2 ratio)
    across 128, 256, and 512 GPUs for LLaMA-3-8B (circles) and Qwen3-32B
    (triangles). Throughput is normalized to \ours, and the vertical 
    dashed line indicates the 600-sec search budget.}
    \label{fig:quality_vs_time}
\end{figure}

Generating plans for distributed training requires joint execution-optimization 
across multiple parallelism dimensions (e.g., data, pipeline, tensor, and
expert) under hardware constraints such as GPU memory limits. However, the
search space for such plans explodes combinatorially with each axis of
heterogeneity, including GPU type, node size, and interconnect bandwidth.
Early planners like Alpa~\cite{alpa} and AMP~\cite{amp} sidestep this
complexity by restricting supported heterogeneity. More recent systems such
as Metis~\cite{Metis}, Espresso~\cite{Espresso}, HexiScale~\cite{HexiScale},
and Sailor~\cite{Sailor} broaden coverage to diverse GPU types and node
sizes. Yet they navigate the vast search space via nested decision loops
built on dynamic programming, graph partitioning, or tree search,
and apply domain-specific pruning to
stay tractable (\autoref{tab:planner-comparison}).

This search paradigm exposes an inherent conflict between planning latency
and quality. Because the nested search loops optimize each plan \emph{from
scratch}, they fail to transfer model-to-GPU mapping principles across
cluster configurations. Without such reusable knowledge, each resource shift
forces planners to rerun the full combinatorial search from an empty state,
which can take tens of minutes or even hours. To reduce this overhead,
planners prune the search with rigid constraints,
such as uniform tensor parallelism across pipeline stages or discarding
partial-node GPUs. These rules, however, routinely discard viable,
high-throughput plans, so the plans that survive fall well short of what the
cluster can deliver (\autoref{fig:quality_vs_time}). Planners today are thus
caught in a dilemma:
thorough search fails to keep pace with frequent resource changes, while
aggressive pruning risks converging on highly suboptimal plans. Furthermore,
none of these planners explicitly support expert parallelism, despite its
critical role in frontier models like DeepSeek~\cite{deepseekv3} and
Qwen~\cite{Qwen}.

Our key insight is that a learned planner need not choose every parallelism
parameter. \emph{Structural} decisions---pipeline depth, per-stage GPU type,
and tensor parallelism (TP) degree---determine how heterogeneous resources
are organized into a pipeline. These decisions involve coupled trade-offs
among compute speed, memory capacity, and communication cost. Once this
structure is fixed, the remaining parameters can be resolved efficiently
using hardware profiles: layers are allocated to balance stage times subject
to memory limits, replica counts follow from GPU packing, and micro-batch
sizes and expert layouts are selected through bounded candidate evaluations.
This decomposition lets a policy focus on structural choices. For each
proposed structure, profile-based rules and bounded candidate evaluations
assign the remaining parameters, allowing the resulting plan's throughput
to guide learning.

We present \ours{}, a reinforcement learning (RL)-based planner that realizes
this decomposition through \emph{structural templates}. Each template
specifies a pipeline's depth and each stage's GPU type and TP degree. The
policy constructs templates sequentially, and a profile-based cost model
completes and evaluates each proposal. Offline training amortizes structural exploration
across GPU compositions, enabling fast runtime planning for unseen cluster
sizes, type mixes, and node sizes of the profiled GPU types without
retraining. The same decomposition accommodates expert parallelism through
bounded expert-layout evaluations without expanding the learned action
space, enabling \ours{} to plan all four parallelism dimensions jointly.
To turn fast re-planning into fast adaptation, \ours{} also integrates a
peer-to-peer GPU state migration mechanism that transfers model state
directly across plan transitions without remote storage.

We implement \ours{} on top of Megatron-DeepSpeed~\cite{megatron-lm,
deepspeed} and compare it against five state-of-the-art planners across
diverse heterogeneous setups (\autoref{tab:planner-comparison}). Each policy
is trained offline and evaluated on unseen GPU compositions without
retraining. In the main dense-model experiments, \ours{} plans clusters of
up to 512 GPUs in 0.7--8.4 seconds, achieving up to 1.85$\times$ the throughput
of the best baseline, while broader-search baselines often exhaust their
600-second budget. On MoE models, its structural choices deliver up to
4.6$\times$ the throughput of the best baseline, and expert parallelism
additionally enables feasible plans for a DeepSeek-V3-like model that cannot
fit under the baselines' expert layouts. On real hardware, \ours{} sustains
19.9\%--69.1\% higher throughput than the best baseline as GPUs join a running
job, demonstrating that fast planning translates into effective adaptation.

%% file: 02_background.tex
\section{Background}
\label{sec:background}

Distributed training
planners typically compose 3D parallelism that combines data (DP), tensor
(TP), and pipeline parallelism (PP). DP replicates model weights across
workers, each processing a distinct slice of the global batch, and
synchronizes gradients every training step~\cite{switchML}. TP partitions
weight matrices within a layer across devices, requiring high-bandwidth
interconnects (e.g., NVLink) due to frequent all-reduce
collectives~\cite{megatron-lm}. PP partitions model layers into sequential
stages hosted on different devices and divides a per-replica batch into
micro-batches to overlap execution and mitigate pipeline bubbles~\cite{gpipe,
pipedream, varuna}.

\PP{Expert parallelism} Frontier models such as Mixtral~\cite{mixtral} and
Qwen3~\cite{Qwen} replace dense feed-forward blocks with Mixture-of-Experts
(MoE) layers, each comprising $E$ experts and a router that dispatches each
token to $k$ of them~\cite{gshard, switchformer}. This design makes the capacity of the model
independent of per-token compute by activating only $k$ experts.
Thus, modern MoE models scale $E$ into the hundreds, causing aggregate
expert weights to dominate the overall memory footprint~\cite{deepspeedmoe,
deepseekv3}.

\begin{figure}[t]
    \centering
    \includegraphics[width=0.46\textwidth]{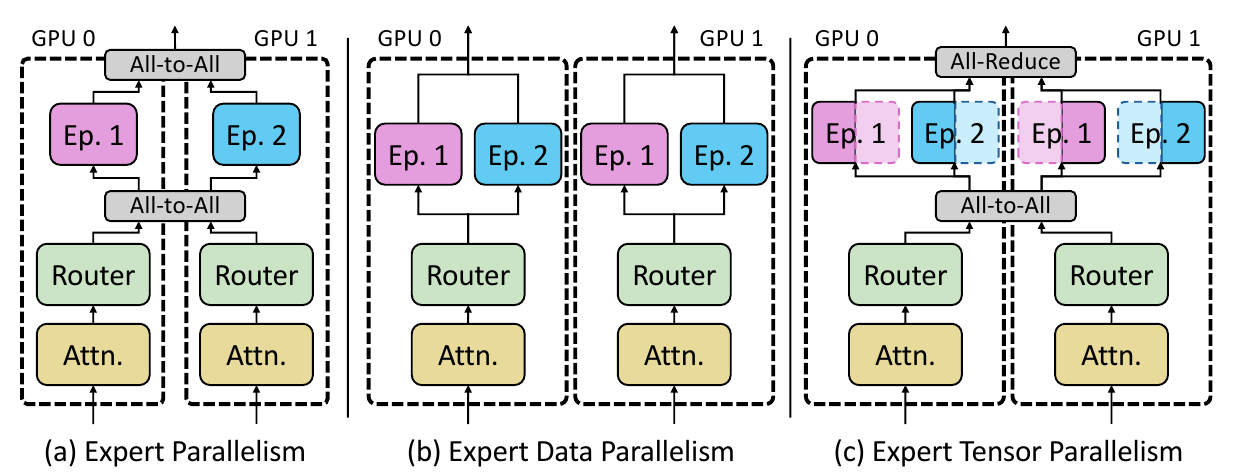}
    \caption{Parallelism strategies for MoE architecture.}
    \label{fig:moep}
    \vspace{-3mm}
\end{figure}

Expert parallelism (EP) scales MoE layers by assigning disjoint subsets of
experts to different devices. It routes tokens via two all-to-all collectives
(\autoref{fig:moep}(a)): one for dispatching tokens to experts and another
for combining their outputs~\cite{megascalemoe}. Similar to TP's all-reduce,
the cost of these collectives remains modest within a node equipped with
high-speed links, but can increase significantly when an EP group crosses
node boundaries without effective communication-computation overlap. Because
an expert block immediately follows an attention block (often parallelized
with TP) in each layer, systems commonly co-locate both on the same GPUs
within a node to ensure that both intra- and inter-block collectives fully
exploit high-speed interconnects. Nonetheless, we can apply 
TP and DP on experts to form expert tensor parallelism (ETP; \autoref{fig:moep}(c)),
which parallelizes individual expert execution typically over intra-node
GPUs, and expert data parallelism (EDP; \autoref{fig:moep}(b)), which
replicates experts across data-parallel groups~\cite{deepspeedmoe,
megascalemoe}. Modern systems like Megatron-Core therefore decouple attention and MoE
parallelism, allowing each block to independently adopt its optimal
configuration on the allocated GPUs~\cite{moefolding,tutel}.

\begin{figure}[t]
    \centering
    \includegraphics[width=0.46\textwidth]{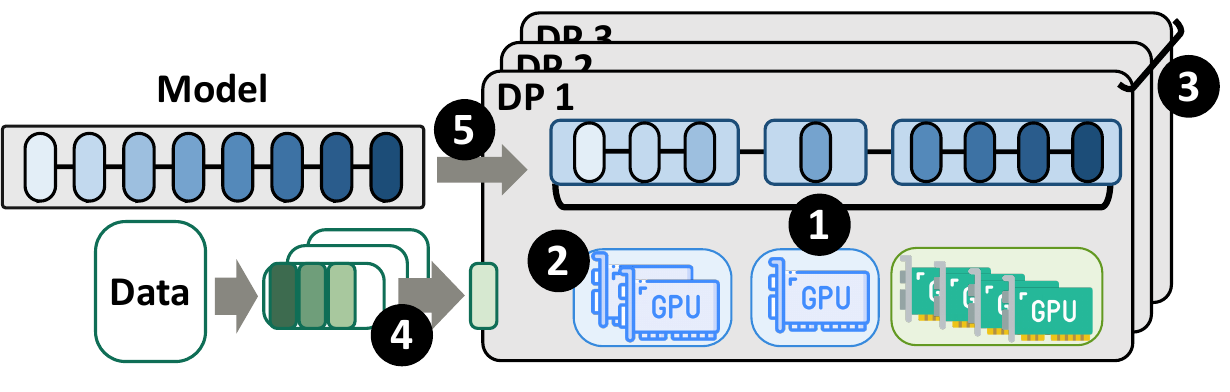}
    \caption{Decision knobs on 3D parallelism planning.}
    \label{fig:solver_component}        
    \vspace{-3mm}
\end{figure}

\PP{Planning under GPU heterogeneity} Prior work for heterogeneous
clusters~\cite{amp, Metis, HexiScale, Sailor, Espresso} tackles the planning
problem solely within the 3D space (DP, TP, and PP) for dense models, with
the primary objective of maximizing throughput (minimizing training step
time) subject to per-GPU memory constraints. For a given model and cluster,
these planners mainly search across five key knobs, as shown
in~\autoref{fig:solver_component}: \ding{182} PP degree (number of stages),
\ding{183} per-stage GPU type and TP degree, \ding{184} DP degree (pipeline
replica count), \ding{185} micro-batch size (MBS), and \ding{186} layer
distribution across stages.

Heterogeneous GPU compositions introduce significant asymmetries that
complicate this planning process~\cite{amp, Metis, HexiScale, whale}. First,
TP requires all participating devices to hold equal tensor slices, yet
distinct GPU models differ in compute power, memory capacity, and
interconnect bandwidth. This forces the planner to choose a TP degree
tailored to each GPU type. Second, PP relies on balanced stage execution
times to minimize pipeline bubbles.
In a heterogeneous cluster, this
precludes uniform layer partitioning, forcing the planner to explore
asymmetric layer assignments that match the compute capacity of the GPUs
assigned to each stage. Finally, DP synchronizes gradients across replicas,
but replicas spanning different GPU types may finish at different rates and
create synchronization bottlenecks. As a result, the search space undergoes a
combinatorial explosion. For $N$ GPU types with $K$ candidate TP degrees and
up to $D$ pipeline stages, \ding{183} alone yields $(N \times K)^D$
configurations. This exceeds 43 million for a modest setup of $N=3$, $K=3$,
and $D=8$, before even accounting for \ding{184}--\ding{186}.

Despite emerging optimizations for MoE on heterogeneous GPUs~\cite{hetermoe,
hexamoe}, current efforts mostly focus on fixed, hand-crafted parallel
layouts (e.g., placing attention and experts across different GPU
generations) or local rebalancing (e.g., redistributing tokens or experts to
alleviate stragglers). None proposes an automated planner to explore the full
4D configuration space (DP, TP, PP, and EP).

%% file: 03_analysis.tex
\section{Analysis of Existing Planners}
\label{sec:motivation}
\label{sec:mot:enum}

Several systems tackle the problem raised in \autoref{sec:background} using
distinct search strategies~\cite{amp, Sailor, Metis, HexiScale, Espresso}.
\autoref{tab:planner-comparison} summarizes their key structural capabilities.
The five planners in \autoref{tab:planner-comparison} adopt two design choices.
First, they search the five knobs of \autoref{fig:solver_component} jointly.
Second, they prune the resulting space with planner-specific heuristics.
The two choices pull against each other.
Broader coverage of the search space costs planning time, and heavier pruning
risks discarding the high-throughput plans.
This section examines what these two choices cost, through an analysis of the
five planners on a concrete running example.

\subsection{Limitations of Search-Based Planning}
\label{sec:bg:obs2}

\noindent\begin{minipage}{\columnwidth}
\begin{lstlisting}[style=code_snippet, label={lst:nested_enum}]
for pp in {1..D_max}:
 for g[1..pp] in GPU_types:
   for tp[1..pp] in TP_degrees:
     for dp in DP_candidates:
      for mbs in MBS_candidates:
        partition = PartitionLayers(pp, ...)
        plan = (pp, g[1..pp], tp[1..pp],
                dp, mbs, partition)
        throughput, oom = Evaluate(plan)
\end{lstlisting}
\end{minipage}

The nested-loop traversal above illustrates the common search backbone shared
by existing planners. Each planner instantiates this backbone differently,
varying both loop ordering and pruning heuristics to manage search latency:

\begin{myitemize2}
  \item \textbf{AMP}~\cite{amp} enumerates TP, DP, and MBS first, partitions
      layers into stages via dynamic programming based on per-layer costs averaged
      across all GPU types, and subsequently completes stage-to-GPU assignments.
  \item \textbf{Espresso}~\cite{Espresso} does not search the TP degree, so
      enforces a uniform TP degree
      across all pipeline stages and prunes stage-to-GPU mappings via bounded
      tree search.
  \item \textbf{HexiScale}~\cite{HexiScale} first partitions the cluster into
      disjoint data-parallel replicas via bandwidth-based graph partitioning, and
      then independently searches for TP and PP within each replica
      strictly constrained by its assigned GPU pool.
  \item \textbf{Metis}~\cite{Metis} jointly searches stage-to-GPU mappings and MBS
      in the outer loop, while delegating layer allocation to a one-shot heuristic
      and favoring stage replication (DP) over TP for multi-GPU stages.
  \item \textbf{Sailor}~\cite{Sailor} enumerates PP and MBS but
      not TP. For each MBS it computes one TP degree per stage.
      The degree is the larger of the smallest that fits the stage in memory
      and one set by how the stage's time scales with TP. The layer allocation
      is uniform across stages.
\end{myitemize2}

\noindent Because these planners prune nested search loops using their own
heuristic rules, each confines its search to a distinct subspace, which
produces qualitatively different plans for identical cluster setups. We next
empirically demonstrate where each planner falls short due to search-space
pruning.

\begin{figure}[t]
    \centering
    \includegraphics[width=\linewidth]{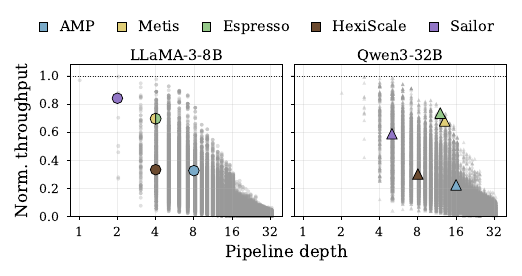}
    \caption{Pipeline structures on the 512-GPU composition of 128
    A100-40, 128 A6000, and 256 H100.
    Gray points are 200,000 samples from the million parallelism plan structures
    drawn at random, scored by the same cost model with baselines
    and the colored points are the five planners' plans.
    Throughput is relative to the best plan found for that model.
    HexiScale is marked on the median pipeline depth.}
    \label{fig:structure_sweep}
\end{figure}

\PP{Setup} We evaluate the five planners on LLaMA-3-8B (32 layers) and
Qwen3-32B (64 layers) across three cluster scales: 128, 256, and 512 GPUs
with a fixed 1:1:2 ratio of A100-40GB, A6000, and H100 GPUs.
\autoref{fig:quality_vs_time} compares their planning quality and latency
under a 600-sec search budget, and \autoref{fig:structure_sweep} shows where
each plan lands within the feasible pipeline search space for the 512-GPU
cluster.

\PP{Observation 1. No existing planner consistently wins across models}
Evaluating the planners on the 512-GPU cluster reveals that they settle on
widely divergent plans in pipeline depth, GPU-type allocation, and stage-wise
TP degrees. No single planner dominates across both models, and all
frequently miss search regions that harbor optimal plans
(\autoref{fig:structure_sweep}). Key issues with each planner include:

\begin{myitemize2}
  \item \textbf{AMP} spreads the model over all three GPU types in eight
  stages at $\text{TP} = 2$. Because its layer partitioning ignores
  hardware speed disparities, stages on slower A6000 and A100-40 GPUs straggle
  behind those on H100s.

  \item \textbf{Espresso} restricts $\text{TP} = 1$ across all stages,
  placing a hard ceiling on throughput.
  It resolves the limitation via a deep $\text{PP}$ degree with layer partitioning.
  On Qwen3-32B, it plans the best plan among baselines. However, the best candidates of
  \autoref{fig:structure_sweep} leverage a TP
  degree, which Espresso cannot consider.

  \item \textbf{HexiScale} partitions the cluster by bandwidth into
  around 100 asymmetric data-parallel pipelines of varying depths. Here, the
  slowest pipeline inevitably bottlenecks cluster-wide gradient
  synchronization.

  \item \textbf{Metis} consistently exhausts its 600-sec search budget due to
  costly enumeration steps. This premature termination leaves exploration incomplete,
  compounding the penalties of search-space pruning.

  \item \textbf{Sailor} achieves competitive throughput on LLaMA-3-8B.
  However, the rigid even-layer partitioning and MBS-TP degree pairing
  lead to a suboptimal decision.
  On Qwen3-32B, it adds the A100-40 GPUs to its H100 pipeline but degrades
  throughput due to its rigid even-layer split.
\end{myitemize2}

\PP{Observation 2. Planning overhead escalates with cluster scale}
As cluster scale and heterogeneity grow, planners
exploring broader search spaces incur prohibitive planning overhead.
\autoref{fig:quality_vs_time} shows that Metis exhausts its 600-sec budget
across all evaluated scales (128, 256, and 512 GPUs). Similarly, Sailor's
search time on LLaMA-3-8B climbs from 106 sec at 128 GPUs to the 600-sec
cutoff at larger scales. Without a time budget, Sailor's search time
reportedly surges from 0.3 sec to 6.2 sec and 4,900 sec as the cluster
expands from one to two and three GPU types (256 GPUs each). Metis takes
hours even with two types~\cite{Sailor}. While HexiScale prunes aggressively
to stay fast, its planning time still grows non-trivially without ever
finding competitive plans.

As in Observation~1, no single planner consistently outperforms the others
across cluster scales. For instance, on LLaMA-3-8B, Sailor ranks last among
the five at 128 GPUs but first at 512. Meanwhile, although Espresso maintains
consistent relative throughput for Qwen3-32B, its plan quality degrades as
the GPU count increases on LLaMA-3-8B.

Together, these observations reveal that existing planners fail to balance
search tractability and plan quality across diverse models and cluster
scales. This limitation is particularly acute in dynamic cloud environments
where GPU availability fluctuates over short time windows~\cite{gandiva-fair,
gavel, MLaaS, tenplex}. In such settings, ML practitioners face an untenable
choice: either settle for highly suboptimal execution under rigid heuristics
or leave newly provisioned GPUs idle while heavyweight planners struggle to
complete in time.


\subsection{Support for Parallelizing Experts}
\label{sec:bg:moe}

Modern MoE systems~\cite{moefolding}
organize the expert layout of a pipeline stage by factorizing it into expert
parallelism ($\mathrm{EP}$), expert tensor parallelism ($\mathrm{ETP}$), and
expert data parallelism ($\mathrm{EDP}$), as introduced
in~\autoref{sec:background}. For a pipeline stage $s$ assigned to $W_s$ GPUs,
the product of their degrees is constrained to $W_s$, i.e.,
$|\text{EP}| \times |\text{ETP}| \times |\text{EDP}| = W_s$. Existing
planners have yet to treat experts as an independent
parallelization knob. Instead, they handle experts identically to dense
attention layers, thus enforcing a uniform stage-wise TP degree that
restricts experts to
$|\text{ETP}| = W_s$ and $|\text{EP}| = |\text{EDP}| = 1$.

This restriction has two consequences. First, planners usually keep a TP
group within a node to run all-reduce over high-speed interconnect,
which imposes  $|\text{ETP}| = |\text{TP}|$.
This leaves no valid assignment when a layer's experts exceed a single node's
memory capacity. Second, even if ETP
can be extended across nodes, doing so severely compromises training
throughput. Slicing experts solely via ETP fragments matrix multiplications
into low-arithmetic-intensity operations and incurs expensive all-reduce
communication. Prior studies report that
relying on ETP alone can drive all-reduce communication cost to 34\% of MoE layer
execution time at 16~GPUs and 57\% at 256~GPUs~\cite{tutel}.
Rebalancing parallelism degrees removes much of this overhead. We observe a 71\% communication time reduction when merely shifting an
eight-GPU stage over two TCP-connected nodes from $|\text{ETP}| = 8$ to
$|\text{EP}| = 8$. In practice, the best configuration varies with GPU memory
capacity and network bandwidth~\cite{hetermoe}. A static, uniform layout
therefore fails to maintain optimal efficiency across a wide range of
heterogeneous resources (\autoref{sec:eval_divec}).

%% file: 04_ideas.tex
\section{Our Approach: Learned Planner}

We first motivate our choice of a reinforcement learning (RL)-based planner
and then present key ideas that make it practical in our target scenarios.

\subsection{Design Rationale}
\label{sec:bg:learn}

With expert parallelism, conventional nested-loop search faces a severe
combinatorial explosion. This makes it far more difficult to design a
search-based planner that identifies optimal plans within practical latency
bounds. Heuristic pruning cannot rescue nested-loop search amid such
complexity, as hard-coding manual rules into a rigid search loop is
inherently brittle. In particular, planning knobs are tightly intertwined,
resource pools cannot be predicted \textit{a priori}, and hardware shifts
quickly invalidate static logic tuned to earlier operating regimes
(\autoref{sec:bg:obs2}).

To overcome these limitations, we consider a learning-based planner that
learns structural decisions and resolves the remaining parameters through
profile-based rules and bounded evaluations. This decomposition lets an RL
agent construct pipeline structures autoregressively and receive throughput
feedback for each completed proposal. Training the policy offline across
diverse GPU compositions amortizes structural exploration, so it can generate
high-throughput plans promptly when a cluster reconfigures. The policy learns
from simulation without costly trial runs on physical GPUs.

\PP{Prior learned approaches operate at an incompatible granularity} Existing
learned frameworks are not suited to our problem: they operate at the
operator level rather than the pipeline level. Prior RL approaches map
individual graph operators to small homogeneous clusters of 2 to 8
GPUs~\cite{hierarchicalRL, placeto, HSDAG}. At the scale of hundreds of
heterogeneous GPUs, this fine-grained search space explodes with both model
graph size and cluster scale and does not explicitly capture macro-level
structures, such as pipeline depth and DP replica counts. We thus need a
planner that operates at a structural abstraction beyond individual
operators.

\PP{Directly learning the raw plan space is sample-inefficient} Training a
policy over all planning knobs in~\autoref{fig:solver_component} faces an
expansive action space where invalid configurations are pervasive. This is
best illustrated by memory heterogeneity across GPU types: it causes many
sampled knob combinations to fail with out-of-memory (OOM) errors.
Consequently, the policy receives virtually no performance feedback to guide
learning while wasting its exploration budget on dead-end configurations. 
We empirically confirm that training directly over the raw knobs needs
2$\times$ more plan evaluations than our approach to reach the same plan
(\autoref{sec:design_validation}).
The central design challenge for a
practical learned planner is therefore not whether learning is viable, but
\textit{which} decisions genuinely require learning and how the remaining
decisions should be resolved.

\subsection{Key Ideas}
\label{sec:ideas}

\autoref{sec:motivation} demonstrates that search-based planners solve each
GPU composition from scratch without transferring knowledge across
invocations. A learned planner amortizes this combinatorial search cost
during offline training and reuses structural policies at runtime. Realizing
such a planner over heterogeneous clusters introduces three key requirements.
\begin{enumerate}
\item The planner must express all four parallelism dimensions (DP, PP, TP, and EP). %
\item The decision formulation must scale to hundreds of heterogeneous GPUs rather than a handful of devices. %
\item The policy must generalize across diverse GPU compositions without retraining.%
\end{enumerate}

\PP{Decomposing monolithic decision spaces} 
The parallelism knobs influence one another, which is why a planner searches
them together. The influence runs both ways, yet the decisions can still be 
resolved in an order.
The pipeline structure decides which devices communicate and over which
links, and the remaining values are bounded once that structure is fixed.
Decisions regarding pipeline stage placement and tensor parallelism directly
interact with cluster topology, creating discrete and non-linear
communication trade-offs across asymmetric interconnects. Conversely,
parameters such as batch sizes and replica counts primarily manage local
memory budgets and scaling efficiency, which behave more predictably once
device assignments are known. The layer share of a stage and the expert
layout of an MoE stage fall on the same side. A stage with a fixed GPU type
and TP degree bounds both, and the all-to-all of an expert layer stays inside
it (\autoref{sec:bg:moe}). This distinction suggests a practical decoupling.
A planner can direct its optimization effort toward the topology-sensitive
placement decisions while resolving the remaining operational parameters
through rule-based or bounded evaluations.

\PP{Make every construction step informative} The split shrinks what the policy
explores, but infeasible choices remain in it.
A depth can exceed the GPUs that remain, a TP degree can exceed its node, and
a stage can still run out of memory.
Two properties therefore have to hold of the construction.
The policy must never sample a choice that cannot run, so that exploration
spends its budget on plans that return a throughput.
Every step must also be judged by a complete plan, so that a partial
construction still returns a signal.
\autoref{sec:design} details how \ours{} realizes these ideas and the three
requirements above.

%% file: 05_design.tex
\section{\ours{}}
\label{sec:design}

\begin{figure}[t]
    \centering
    \includegraphics[width=0.46\textwidth]{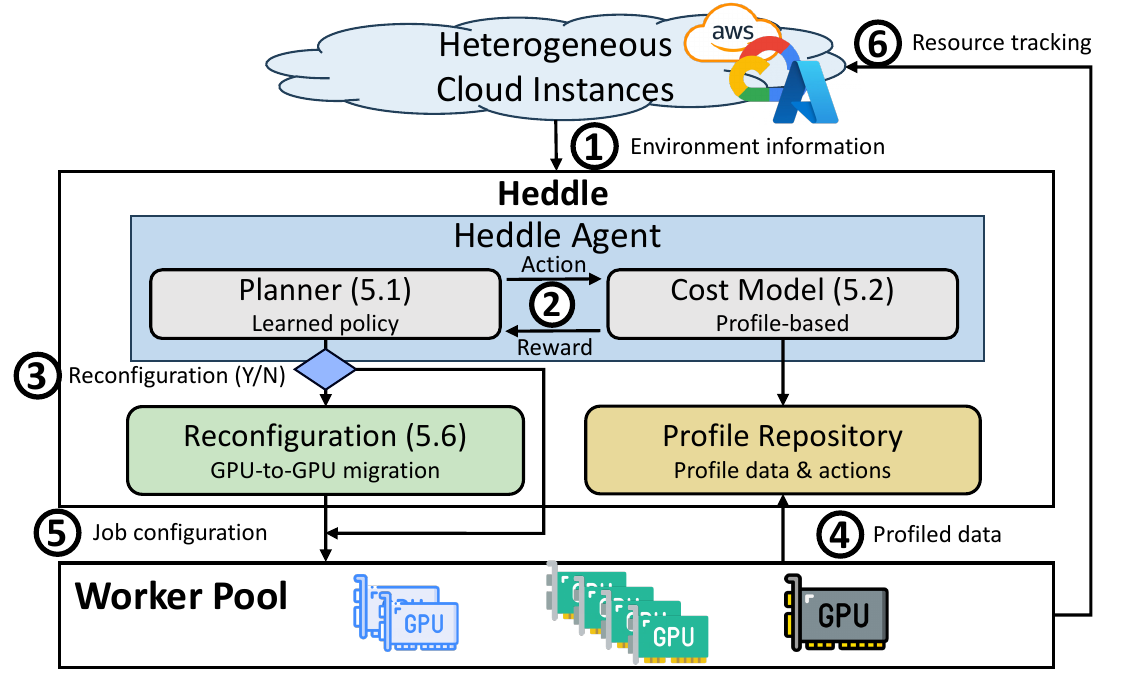}
    \caption{\ours{} system overview.}
    \label{fig:overivew_diagram}
    \vspace{-3mm}
\end{figure}

Building on the design principles in~\autoref{sec:ideas}, \ours{} decouples
parallelism planning around a central abstraction called the
\textbf{structural template}, which encapsulates
structural decisions into reusable units. To operate atop this abstraction,
\ours{} coordinates three core components: (1) a \emph{learned planner}
(\autoref{sec:planner}) for autoregressive search over structural templates,
(2) a \emph{profile-based cost model} (\autoref{sec:simulator}) for resolving
arithmetic dimensions (\autoref{sec:fill}) and checking execution feasibility, and (3) a
\emph{reconfiguration module} (\autoref{sec:Reconfiguration}) for state
migration during plan transitions.

\autoref{fig:overivew_diagram} illustrates how these components interact
end-to-end. Given a target GPU composition~\ding{192}, the planner proposes
candidate structural templates via an offline-trained policy~\ding{193}. For
each template, the cost model auto-fills arithmetic parameters (e.g., DP
degree, layer distribution, micro-batch size, and expert tiling) and ranks
the resulting plans by predicted throughput using offline
profiles~\ding{195}. If training is starting from scratch~\ding{194}, the
highest-throughput plan is deployed directly to workers~\ding{196}. When
re-planning is triggered by a resource shift during training~\ding{197}, the
reconfiguration module instead migrates model state between GPUs to adopt the
new plan without remote storage~\ding{196}.

\subsection{Learning-based Planner}
\label{sec:planner}

\PP{Structural template as the primary unit of planning} A structural
template defines the pipeline topology by assigning a GPU type and a TP
degree to each pipeline stage (\ding{182} to \ding{183}
in~\autoref{fig:solver_component}). All devices within a stage share these
attributes, establishing the stage as the unit of resource allocation. Once a
template is fixed, arithmetic decisions are resolved post-hoc via rule-based
heuristics and bounded profile lookups (\ding{184} to \ding{186}
in~\autoref{fig:solver_component}), as detailed in~\autoref{sec:fill}.

This template formulation satisfies the three operational requirements
outlined in~\autoref{sec:ideas}. First, the template configuration,
replication factor, and stage-wise expert sizing jointly express all four
parallelism dimensions (DP, PP, TP, and EP). 
Second, the structural action
space depends solely on the number of GPU types, TP options, and pipeline
depth, so it does not grow with the layer count of the model.
Third, the state describes a cluster by the fraction and the device
properties of each GPU type, and by the cluster scale.
One policy therefore transfers to new sizes and node mixes of the GPU types
it trained on, without retraining.

To transfer across cluster scales, each GPU type is characterized by
normalized hardware attributes such as peak compute throughput (TFLOPS),
memory capacity, and speed at each TP degree. Meanwhile, the cluster state
records the relative proportion of each GPU type and the cluster scale
instead of absolute per-device counts. The cluster's scale itself is also in the
state, so the policy is not bound to that ratio. It can choose a deeper pipeline
where a larger cluster favors a different structure. 
For example, when a cluster scales from 8~A100s and 16~H100s to 32~A100s and
64~H100s, the 1:2 A100-to-H100 ratio is preserved while the scale increases.
Through this representation, learned placement policies transfer directly across cluster scales.

\begin{figure}[t]
    \centering
    \includegraphics[width=0.46\textwidth]{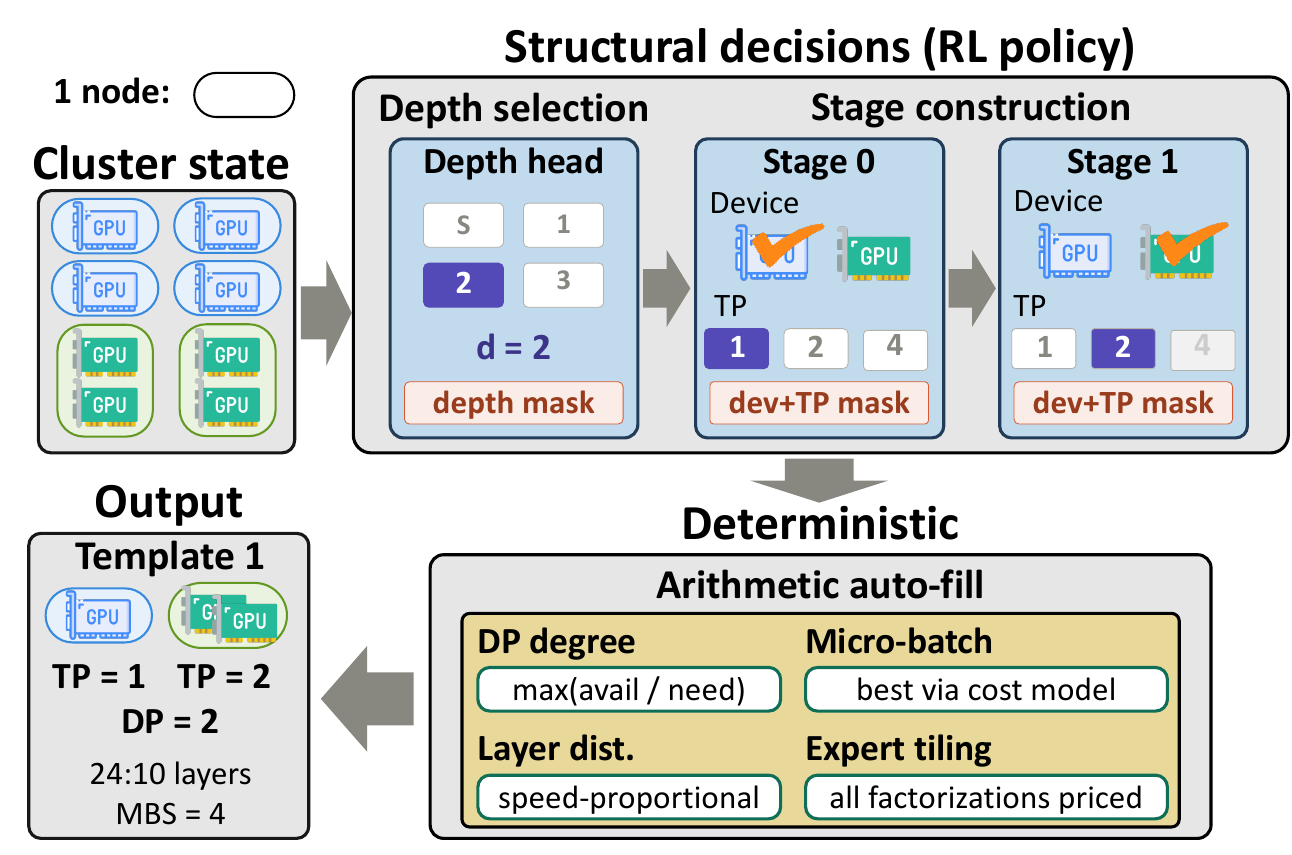}
    \caption{\ours{} template building process.
    At each construction step, the planner selects a
    pipeline depth, then autoregressively assigns a
    GPU type and TP degree to each stage.
    The system auto-fills the layer distribution, DP degree,
    micro-batch size, and the expert tiling of an MoE model.
    The process repeats until the planner emits
    \texttt{STOP} or the GPU composition is fully mapped.}
    \label{fig:template_building}
\end{figure}

\PP{Planning as autoregressive construction}
\ours{} constructs a plan as a sequence of structural templates
(\autoref{fig:template_building}).
Each step adds one template, and the construction ends when the planner emits
a \texttt{STOP} signal or reaches the step limit.
A plan is the set of templates placed when construction stops, and each of
them holds its own GPUs.
A step first selects a pipeline depth $d \in \{1, \dots, D_{\max}\}$.
It then assigns a GPU type and a TP degree to each of the $d$ stages, one
stage at a time.
The TP degree comes from $K = \{1, 2, \dots, 2^{n}\}$, where $n$ follows the
maximum number of GPUs in a node, so $K = \{1, 2, 4\}$ for 4-GPU nodes.
\ours{} fills the arithmetic dimensions of a template as soon as a step
completes it.

\PP{Feasibility masking}
A mask removes the choices that cannot run before the planner samples a
decision.
The depth mask leaves out a depth that the remaining GPUs cannot hold.
The device mask leaves out a GPU type with no GPU left.
The TP mask leaves out a degree above the node size of the selected type.
The expert tiling needs no mask, since the fill prices it rather than the
policy choosing it.
A mask changes no relative order among the choices it keeps, so the policy
ranks the feasible ones as it learned to.
Every sampled action therefore completes into a plan the cost model prices,
which is what a sparse reward demands of the planner (\autoref{sec:ideas}).

\PP{Non-uniform GPU composition support}
\ours{} reads a GPU composition as a node list per GPU type, and a node can
hold any number of GPUs~\cite{Metis}.
The TP mask therefore limits a TP degree to the largest node of its GPU type.
A TP group never crosses a node boundary, where the communication would be
much slower.
\ours{} first looks for a node whose size equals the requested TP degree.
If no such node remains, it takes the necessary GPUs from a larger node.

\subsection{Profile-based Cost Model}
\label{sec:simulator}

The cost model reports the iteration time and the per-GPU peak memory of a
plan from offline-profiled measurements, so the planner needs no online
profiling.

\PP{Throughput estimation}
The cost model holds the profiles that \ours{} measures offline.
One profile entry names a GPU type, a TP degree, and a micro-batch size, and
it gives the computation cost and the communication cost of one layer.
The profiling cost is low, because \ours{} profiles one layer per repeated
structure, such as one transformer block, and reuses that cost for every
identical layer.
A model with $L$ unique layer types, usually three or four in an LLM, needs
each entry $L$ times.
A new GPU type or a new TP option needs only the affected entries.
The cost model looks up an exact match only, and a configuration the profiles
do not hold leaves the search space.

\PP{Memory validation}
The cost model computes the per-GPU peak memory footprint by summing model
parameters, optimizer states, gradients, and activation memory for the given
layer partition and micro-batch size.
It compares the footprint against the physical memory of each GPU, so it finds
an out-of-memory (OOM) configuration before it estimates the throughput.
\ours{} applies this check during plan construction and during the fill of
\autoref{sec:fill}.

\subsection{Arithmetic Auto-fill}
\label{sec:fill}

\ours{} resolves the arithmetic dimensions of every candidate the policy
proposes, and none of them enters the action space.
Each dimension either follows from the template or comes from a bounded
candidate set that the cost model prices.

\PP{Layer split and DP degree}
The layer split follows from the template.
\ours{} gives each stage layers in proportion to its profiled speed, so the
stages of a pipeline take a similar time.
If that split exceeds the memory of a stage, \ours{} prices a fixed list of
alternative splits and keeps the fastest one that fits.
The DP degree follows from the remaining nodes.
\ours{} packs the TP group of each stage into the nodes of that stage's GPU
type and takes the largest replication count that the packing admits.

\PP{Micro-batch size}
The micro-batch size comes from the profiled powers of two.
\ours{} takes the largest size that fits in memory and every smaller one, and
it keeps the size with the shortest iteration time.
The cost model scores each candidate as part of the plan so far, so the choice
also reflects the templates that \ours{} already placed.

\PP{Expert tiling}
The expert layout of a stage factorizes into EP, ETP, and EDP, whose product
is the number of devices $W_s$ mapped to the stage.
\autoref{sec:bg:moe} gives the cost of each direction.
$\mathrm{ETP}$ carries the node bound of $\mathrm{TP}$, since both slice one
tensor.
$\mathrm{EP}$ does not, and \ours{} prices a group that leaves a node with the
bandwidth measured across nodes rather than the one measured inside.
$\mathrm{EDP}$ is then what remains, $W_s$ over the product of the other two.
\ours{} prices every factorization of $W_s$ that fits in per-GPU memory and
keeps the one with the shortest execution time.
The node size and the device memory leave 3 to 12 such factorizations in
practice. %
A stage can be tiled on its own, because its choice does not reach its
neighbors.
The dispatch and combine collectives of an expert layer stay within the
devices of the stage, so the tiling of stage $s$ changes $t_s$ and the
gradient synchronization $g_s$ of that stage alone.
An iteration of a 1F1B pipeline takes $(M-1)\max_s t_s + \sum_s t_s + \max_s
g_s$ for $M$ micro-batches.
The first two terms grow with every $t_s$, so the fastest tiling for each
stage is also the fastest for them.
\ours{} therefore takes the per-stage choice as a starting point and
re-selects against the whole iteration, over the candidates it has already
enumerated.

\subsection{Policy Network}
\label{sec:state_action}

\PP{Planner state}
The state is a vector of fixed size, so one policy transfers across
compositions without retraining.
Its entries are fractions and normalized device properties, and the cluster
size enters on a log scale.
\autoref{tab:state_features} lists the per-type features and the cluster-wide
features.
The composition group tells apart clusters that share every device property
but differ in mix.
To relate every candidate depth to the current cluster size, the state holds
the GPU count over that depth.
The count is the most replicas a depth can leave when a stage takes one GPU.
The state ends with a short summary of the composition, which gives its size,
the share of each type, and how uneven the mix is.
A small network turns that summary into a scale and a shift for every layer of
the backbone and for every head.
Therefore, the same trained weights act differently on different compositions.

\begin{table}[t]
\centering
\caption{Planner state features.
$K$ is the set of TP degrees, the powers of two up to the node size, and $G$
the GPU count.}
\label{tab:state_features}
\footnotesize

\begin{tabularx}{\columnwidth}{@{}lX@{}}
\toprule
\textbf{Feature} & \textbf{Description} \\
\midrule
$\rho_i$, $\eta_i$ & Remaining GPU and node fractions of type $i$ \\
$s_i$ & Fraction of all GPUs that are of type $i$ \\
$\bar{f}_i$, $\bar{m}_i$ & Normalized peak TFLOPS and memory \\
$\alpha_i^{(k)}$, $\phi_i^{(k)}$ & Relative and effective speed at TP $k \in K$ \\
$\mu_i^{(k)}$ & Layer memory at TP $k$ over GPU memory \\
$h_i^{(k)}$ & Remaining nodes of size $k$ \\
$\log G$ & Cluster scale \\
$t/T_{\max}$ & Construction step progress \\
\bottomrule
\end{tabularx}
\end{table}

\PP{Decision heads}
The policy network emits the decisions of a step through three heads.
The depth head selects the PP degree from $\{\texttt{STOP}, 1, \dots,
D_{\max}\}$. Selecting \texttt{STOP} ends the construction, and selecting a depth starts to build the pipeline structure.
The device head then works through the stages of that depth, one at a time, and
gives each stage a GPU type.
The TP head gives the same stage a TP degree.
Each head reads what the decisions before it have fixed.
This order narrows each decision.
The depth fixes how many stages the step has to fill.
A placed stage takes its GPUs out of what the later stages can use.
A GPU type admits only the TP degrees that fit one of its nodes.
The depth head and the device head score a candidate together with the features
of that candidate, so a logit moves when the cluster changes.

\subsection{Training the Planner}
\label{sec:training}

\ours{} trains one policy per model offline, on generated compositions of the profiled GPU types, and plans compositions it has not seen. Training needs no GPU, because the cost model prices every rollout.

\PP{Reward}
The per-step reward is the throughput delta
\begin{equation}
\label{eq:reward}
  r_t =
  \begin{cases}
    \mathrm{Thr}(P_t) - \mathrm{Thr}(P_{t-1}),
      & \text{if } a_t \text{ is valid}, \\
    -\delta,
      & \text{otherwise (e.g., OOM)},
  \end{cases}
\end{equation}
where $\mathrm{Thr}(P_t)$ is the throughput of the plan built so far, after the fill of \autoref{sec:fill} has completed the new template.
Every step is therefore judged by a complete plan, which is what \autoref{sec:ideas} asks of the signal. 
The term $\delta$ is a fixed penalty for a template that does not fit, and it applies at the first step only. 
At a later step such a template ends the episode at the plan built so far, with no penalty. The delta also teaches the planner when to emit \texttt{STOP}, since a template that lowers throughput earns a negative reward.

\PP{Training one policy across compositions}
\ours{} trains one policy for every composition, so a single update has to
learn from all of them at once.
The reward and the gradient are not comparable across compositions, and
\ours{} makes each comparable before the update uses it.
The reward is a throughput, and throughput differs by an order of magnitude
across compositions and models.
\ours{} therefore keeps the clipped objective of PPO~\cite{PPO} but judges each
rollout against the other rollouts of its group~\cite{grpo}.
A group is a set of rollouts on one composition from the same state, so its
members differ only in the decisions the policy made.
A rollout above the others of its group earns a positive advantage, and one
below them a negative one, whatever the scale of that composition.
\ours{} therefore needs no learned value baseline, which would first have to
learn the scale of every composition.
The gradients of two compositions can pull a shared weight in opposite
directions when one holds a GPU type the other does not have.
When two point against each other, \ours{} removes from each the component that
lies along the other~\cite{pcgrad}.
What is left still improves its own composition, but no longer at the expense
of the other.

\PP{Exploration in training}
Two rules keep deep pipelines in every group, because shallow pipelines dominate early training.
A shallow pipeline gives positive throughput with most GPU-type and TP-degree combinations, so it
can hide a fast deep pipeline. With a probability that decays from 0.8 to 0.25 over training, a
rollout is forced to a feasible depth drawn uniformly. One rollout per group draws its GPU types
uniformly instead of from the policy. An entropy bonus keeps the remaining rollouts from collapsing
onto one template. \ours{} computes it per decision and divides it by the number of decisions in a
step, so that depth alone does not earn a larger bonus. Self-Imitation Learning (SIL)~\cite{Oh2018SIL}
with a replay buffer per depth anchors the planner to the best plan found at each depth. At a fixed
interval the planner imitates the best plans in these buffers and those of similar compositions.

\begin{algorithm}[t]
\caption{Parallelism planning via template
construction.}
\label{alg:template_building}
\begin{algorithmic}[1]
\Require GPU composition $\mathcal{C}$ ($N$ GPU types), model profile $\mathcal{P}$, learned planner $\pi_\theta$, max depth $D_{\max}$, max steps $T_{\max}$, TP options $K$, expert count $E$ ($E{=}1$ for dense)
\Ensure Parallelism plan
\State $s \leftarrow \textsc{Reset}(\mathcal{C}, \mathcal{P})$; \; $\textit{done} \leftarrow \textsc{False}$
\While{not \textit{done}}
    \State $m_d \leftarrow \textsc{DepthMask}(s)$
    \State $d \sim \pi_\theta.\textsc{DepthHead}(s, m_d)$
        \Comment{Select depth or \texttt{STOP}}
    \If{$d = \texttt{STOP}$}
        \State $\textit{done} \leftarrow \textsc{True}$; \textbf{continue}
    \EndIf
    \For{$j = 0, \dots, d{-}1$}
        \Comment{Stage construction}
        \State $g_j \sim \pi_\theta.\textsc{DevHead}(s, d, j, \textsc{DevMask}(s))$
        \State $\tau_j \sim \pi_\theta.\textsc{TPHead}(s, d, j, g_j, \textsc{TPMask}(s, g_j))$
    \EndFor
    \Statex \hspace{\algorithmicindent}\textit{// Arithmetic auto-fill (\autoref{sec:fill})}
    \State $\ell \leftarrow \textsc{Distribute}(\text{template}, \mathcal{P})$
    \State $n, b \leftarrow \textsc{CostModel.BestFill}(\text{template}, \ell, s)$
    \State $e \leftarrow \textsc{CostModel.BestTiling}(\text{template}, E)$
    \State $s, \textit{done} \leftarrow \textsc{Step}(\text{template}, \ell, n, b, e)$
\EndWhile
\State \Return best plan
\end{algorithmic}
\end{algorithm}

\subsection{Reconfiguration}
\label{sec:Reconfiguration}

When the GPU composition changes during training, the model state has to move
to the GPUs of the new plan.
\ours{} moves it GPU to GPU~\cite{Sailor,tenplex,recycle,oobleck} rather than
through remote-storage checkpoints~\cite{gemini-checkpoint,universal_checkpointing}.
Its Direct State Migration (DSM) copies state between two device memories with
a peer-to-peer transfer of NCCL~\cite{hu2025demystifying}.
The difficulty is that NCCL needs a deterministic synchronization.
The two GPUs of a transfer must agree on the data volume and on the peer before
the transfer starts.
DSM meets this condition in two stages, so transfers run in parallel without a
global scheduler.
It first computes the difference between the old and the new plan as a set of
migration tuples.
For each tuple, it picks the sender with the highest-bandwidth link to the
receiver from the profiled topology.
\begin{myitemize}
    \item \textbf{Stage 1 (intra-pipeline).}
    DSM executes the tuples whose sender and receiver lie in one pipeline.
    These transfers are pipeline-local, so every pipeline runs them in parallel
    without mismatches.
    \item \textbf{Stage 2 (inter-pipeline).}
    DSM then executes the tuples that cross pipelines as a sequence of
    point-to-point exchanges.
    The two stages together meet the size and identity conditions of NCCL.
\end{myitemize}

\begin{figure*}[t]
    \centering
    \includegraphics[width=\linewidth]
      {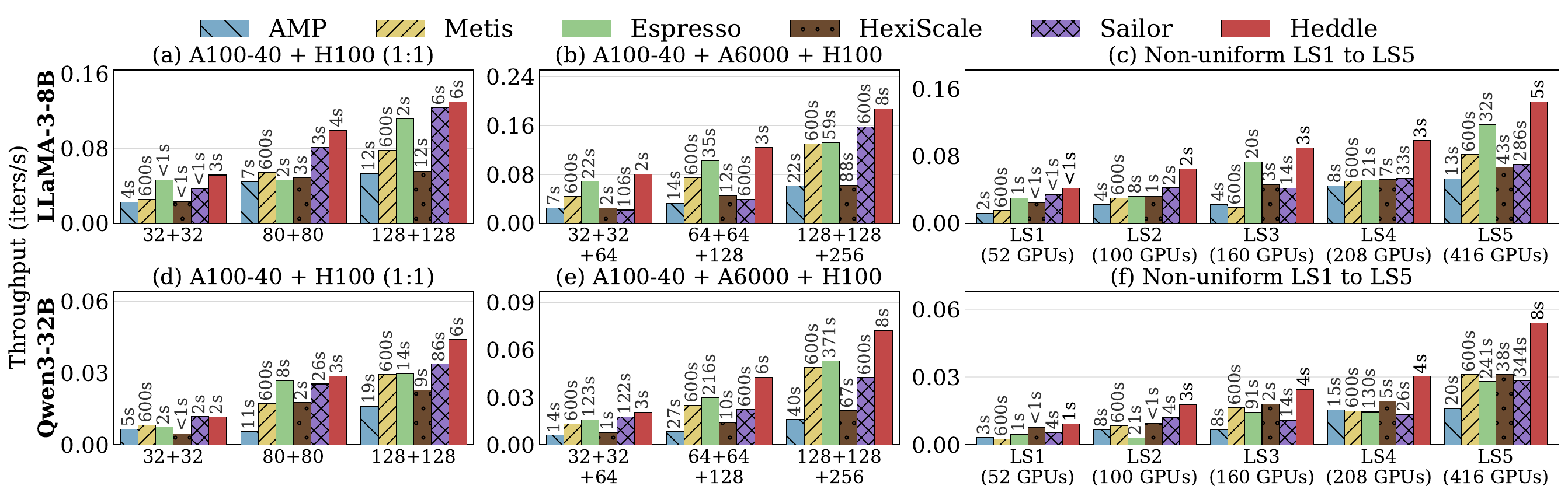}
    \caption{Throughput of the plan each planner returns, for
    LLaMA-3-8B (top) and Qwen3-32B (bottom).
    (a,d) and (b,e) are uniform GPN\,=\,4 compositions of 64 to 512~GPUs
    (x-axis, GPUs per type).
    (c,f) are the non-uniform compositions LS1 to LS5, 
    which contain a heterogeneous mix of A100-40, A100-80, and H100 GPUs 
    across 1-GPU and 4-GPU nodes.
    Numbers above bars are search time in seconds.}
    \label{fig:main_results}
\end{figure*}

%% file: 07_evaluation.tex
\section{Evaluation}
\label{sec:evaluation}

The evaluation answers three questions. (1) How does \ours{} compare with existing
planners on diverse heterogeneous GPU compositions? (2) Does \ours{} extend to diverse dimensions?
(3) How much does each design component contribute to plan quality?

\PP{Models and baselines}
The large-scale simulation study uses LLaMA-3-8B~\cite{llama3}, Qwen3-32B~\cite{Qwen}, and
Qwen3-64B~\cite{Qwen} with 32, 64, and 128 layers, and the hardware validation
uses GPT-Neo-2.7B~\cite{GPT-Neo} with 34 layers on Megatron-DeepSpeed.
All run at sequence length 2048 with FP16 and Adam, at a global batch of 2048
in simulation and 128 on hardware. The baselines are Sailor~\cite{Sailor}, Espresso~\cite{Espresso},
Metis~\cite{Metis}, HexiScale~\cite{HexiScale}, and AMP~\cite{amp}, each
planning with the cost model it was published with, under a 600-sec search budget.
HetAuto~\cite{HetAuto} searches with MCTS and appears only in \autoref{sec:design_validation},
which counts plan evaluations rather than seconds.

\PP{Configurations and training}
The simulated compositions use A100 (40 and 80\,GB), H100, and A6000, where GPN
denotes GPUs per node.
The hardware is five machines with nineteen GPUs of three generations, one node
of three RTX~PRO~6000 (96\,GB), three nodes of four A6000 (48\,GB), and
one node of four A5000 (24\,GB), on 100\,Gb/s Ethernet.
\ours{} trains one policy per model with GRPO.
For LLaMA-3-8B and Qwen3-32B, the policy is trained 1{,}400 episodes over 94
synthetic clusters of two to four GPU types in about two hours. The clusters cover which type holds
the most GPUs, how many types a cluster holds, the cluster size, and the node size.
None of the 94 appears in the evaluation, and that one checkpoint plans every
test composition without retraining.
The hardware policy is trained the same way on 32 clusters of the six GPU types in one hour.

\subsection{Large Scale Simulation}
\label{subsec:perf}

\PP{Uniform GPU composition}
\ours{} returns the highest-throughput plan on eleven of the twelve cells of
\autoref{fig:main_results}(a), (b), (d) and (e), and on the twelfth, Qwen3-32B
with 32 A100-40 and 32 H100, it is within 0.4\% of Sailor.
Each model runs on six compositions of 64 to 512~GPUs.
On Qwen3-32B it leads the closest baseline by 7.5 to 31.0\% on the other two compositions with two GPU
types and by 16.5 to 43.5\% with three types.
On LLaMA-3-8B it leads by 5.1 to 22.9\% with two types and by 17.7 to
21.5\% with three types.
\ours{} plans every composition in 1.6 to 8.4\,s, where Sailor finishes every
two-type composition and then spends its full 600-sec budget on four of the six
three-type ones. Metis reaches its 600-sec budget on all twelve.

\PP{Non-uniform GPU composition}
\autoref{fig:main_results}(c) and (f) repeat the comparison on five
non-uniform compositions, LS1 to LS5, which mix 1-GPU and 4-GPU nodes of three
GPU types at 52 to 416~GPUs.
Metis and HexiScale receive the topology as it is.
Sailor, AMP, and Espresso require uniform node sizes, so each receives the
reading in which every GPU is its own node.
\ours{} can leverage the 1-GPU node with respect to the model size.
\ours{} leads at every size on both models, by 22.4\% to 84.5\% on
LLaMA-3-8B and 21.6\% to 72.9\% on Qwen3-32B.
Against Metis and HexiScale alone the LLaMA-3-8B lead is 71.7\% to 102.3\%.
\ours{} plans each composition in 0.7 to 8.4 sec.

\subsection{Validation on Real Hardware}
\label{sec:realhw}

\begin{figure}[t]
    \centering
    \includegraphics[width=\linewidth]{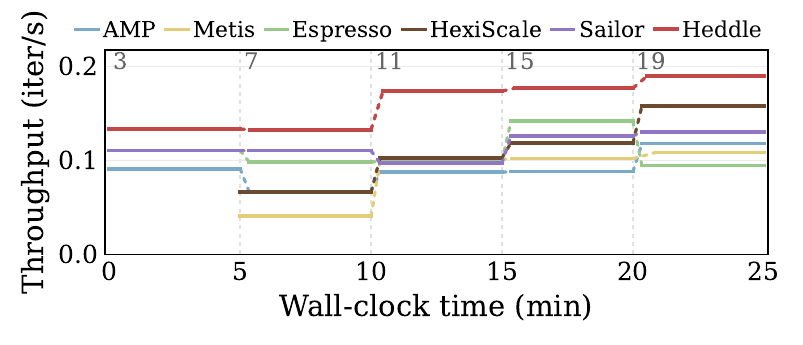}
    \caption{Measured training throughput of each planner while the cluster grows under a running job.
    The number above each segment is the total GPU count after that join.
    A break in a line is the interval for its planning call and for reconfiguration.}
    \label{fig:growth}
\end{figure}

\PP{A cluster that grows under the job}
\ours{} is the only planner whose measured throughput never falls as the
cluster grows.
\autoref{fig:growth} starts the job on the three RTX~PRO~6000 GPUs and adds a
four-GPU node (A6000, A5000, A6000, and A6000 in order) every five minutes, up to nineteen GPUs.
Every planner replans at each join with no plan carried forward, so a plan
slower than the one it replaces appears as a drop.
\ours{} leads at every size by 19.9\% to 69.1\%, where AMP falls by 27\% at
the first join, Sailor by 12\% at eleven GPUs, and Espresso by 33\% at
nineteen.
Increasing nodes does not always guarantee better performance.
\ours{} decides not to change the parallelism plan on the first join and third join.
Also, at the last stage, which has five nodes, \ours{} drops the A5000, which is a straggler.

\PP{Cost model accuracy}
On the testbed the cost model predicts the measured iteration time within 6.1\% on average,
and within 5.2\% for plans that stay inside one machine and 8.8\% across machines.
Inside a single machine, running one plan on real hardware already varies by 2.5\%.

\subsection{Extension to Diverse Dimensions}
\label{sec:eval_divec}

\begin{figure}[t]
    \centering
    \includegraphics[width=\linewidth]{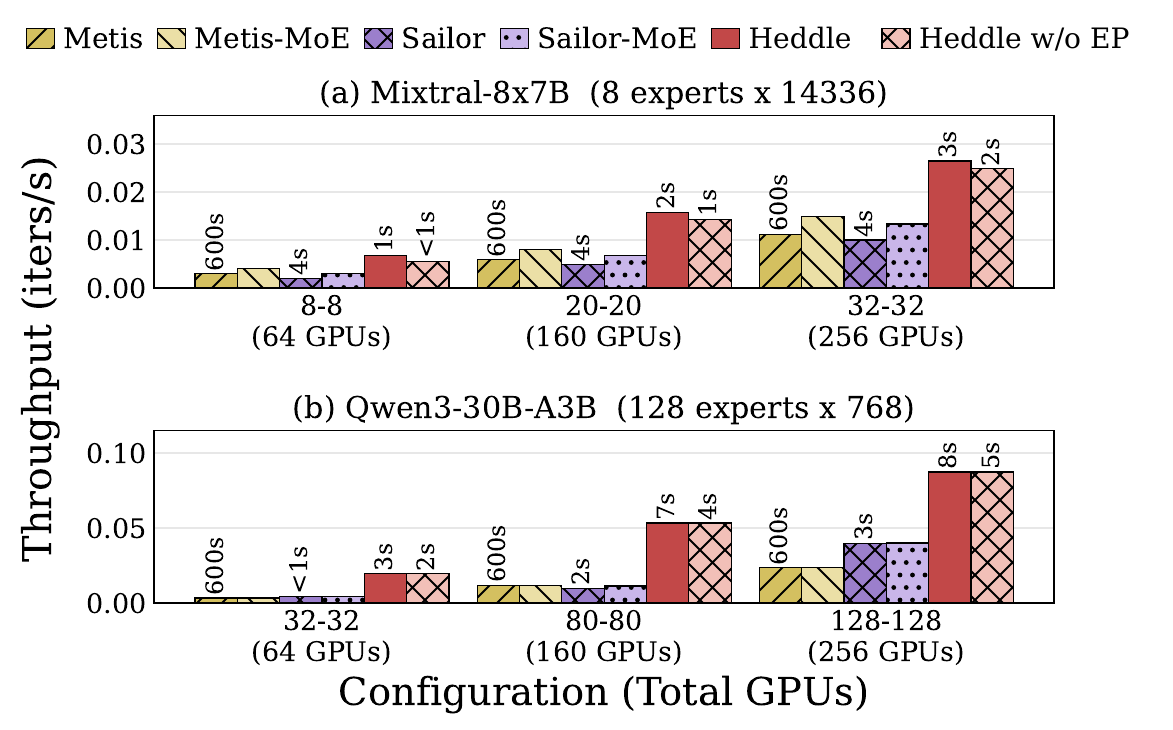}
    \caption{Throughput of the plan each planner returns on two MoE models
    (x axis, total GPU count) with search time in seconds above each bar.
    A \mbox{-MoE} bar re-tiles that planner's own plan by the rule of
    \autoref{sec:fill}.
    \mbox{\ours{} w/o EP} pins every stage to $\mathrm{ETP}=\mathrm{TP}$ and
    $\mathrm{EP}=1$.
    }
    \label{fig:moe}
\end{figure}

\PP{MoE models and the expert parallelism dimension}
What the expert dimension is worth depends on the model, and three models cover the range.
On Mixtral-8x7B it buys the throughput (\autoref{fig:moe}(a)).
\ours{} runs 1.21, 1.10, and 1.06 times faster at 64, 160, and 256~GPUs than the same policy with the dimension without tiling.
Re-tiling the baselines' own plans by the same rule lifts them 33 to 49 \%, and \ours{} still leads the best of them by 1.68 to 1.98 times.
On Qwen3-30B-A3B it makes no difference (\autoref{fig:moe}(b)).
One layer fits on a single GPU whole, so the pricing takes $\mathrm{EP}=1$ on every stage,
and the two \ours{} bars are the same plan and the same number at all three sizes.
\ours{} leads Metis and Sailor by 2.2$\times$ to 4.6$\times$ in this panel.
The lead comes from the plan structure.
On a DeepSeek-V3-like~\cite{deepseekv3} model it buys the plan itself.
One layer holds 168\,GB of expert state, and spreading it over a whole node still leaves more than one GPU can hold (GPN=4).
Sharding experts by the TP degree has nowhere further to go, since a TP group does not leave the node.
Metis, Sailor, and \ours{} with the dimension without tiling therefore return nothing at 160, 256, and 512~GPUs.
\ours{} places it at all three sizes.

\begin{figure}[t]
    \centering
    \includegraphics[width=\linewidth]
    {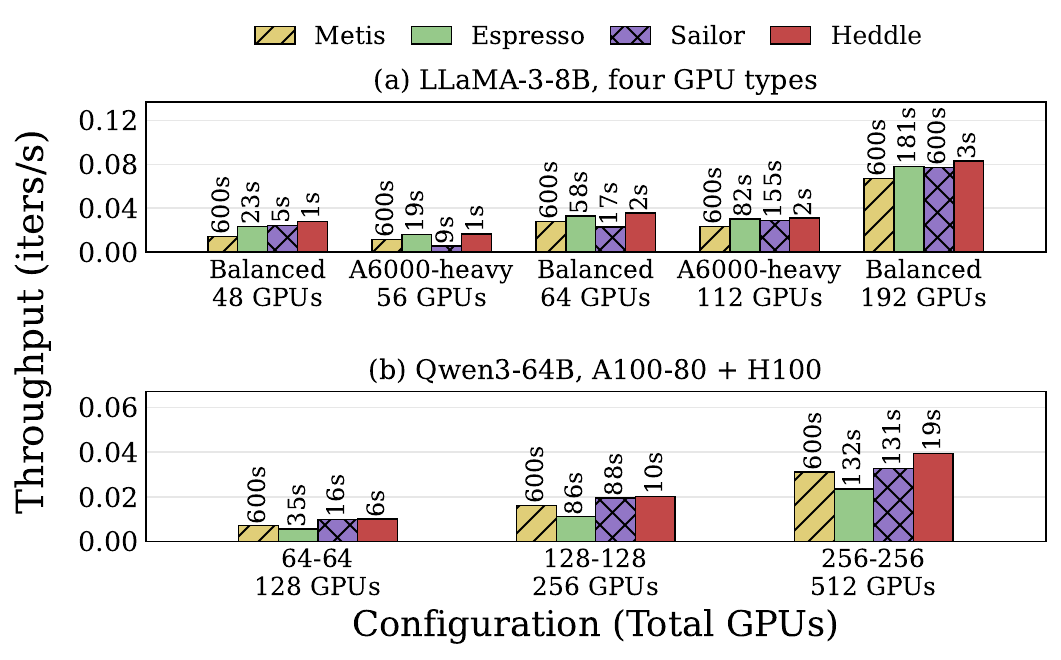}
    \caption{The top panel shows LLaMA-3-8B throughput on four-type
    compositions (A6000, A100-40, A100-80, H100).
    The bottom panel shows Qwen3-64B on A100-80 and H100 GPU composition.
    Numbers above bars are search time in seconds.
    }
    \label{fig:4type_llama}
\end{figure}

\PP{More GPU types}
\autoref{fig:4type_llama}(a) evaluates LLaMA-3-8B on five compositions of four
GPU types, A6000, A100-40, A100-80, and H100, from a balanced 48-GPU cluster to
192~GPUs, two of which give A6000 57\% of the GPUs.
\ours{} returns the highest-throughput plan on all five, ahead of the
best baseline by 1.6\% to 14.3\%.

\PP{Scaling to a larger model}
\autoref{fig:4type_llama}(b) scales the model to Qwen3-64B, Qwen3-32B with its
transformer stack doubled to 128 blocks (62.3\,B parameters), trained with
activation checkpointing on A100-80 and H100 in equal counts at 128, 256,
and 512~GPUs. The policy trained on Qwen3-32B plans it without retraining and 
leads Sailor, the best baseline at all three sizes, by 0.2\%, 3.7\%, and 20.9\%.
Metis spends its budget rejecting 10{,}494 plans as out 
of memory at 128~GPUs and 3{,}159 at 512, where \ours{} plans in 6 sec and 19 sec.

\subsection{Design Validation}
\label{sec:design_validation}

\begin{figure}[t]
    \centering
    \includegraphics[width=\linewidth]{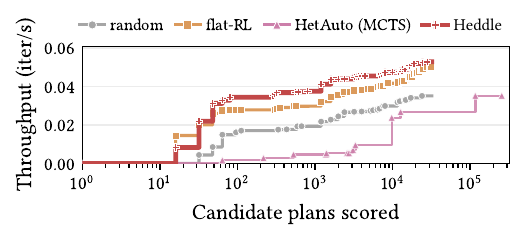}
    \caption{Search efficiency on a 160-GPU composition of four
    GPU types (16 H100, 32 A100-80, 48 A100-40, 64 A6000) for LLaMA-3-8B,
    as best-so-far throughput.
    }
    \label{fig:search_efficiency}
\end{figure}

\PP{Search efficiency}
Plan quality has two sources, and each is worth a different thing
(\autoref{fig:search_efficiency}).
The decomposition buys sample efficiency.
\texttt{flat-RL}, the same policy over the raw six-dimensional action space,
needs twice the plan evaluations \ours{} needs to reach the same plan, because
the three extra dimensions it learns are the arithmetic ones the structure
already determines (\autoref{sec:ideas}), and it stays 5\% to 29\% below
\ours{} at every budget past the first few dozen evaluations.
The offline learning buys direction.
At 32{,}000 evaluations \texttt{random} reaches 0.035~iter/s where \ours{}
reaches 0.053, and \texttt{HetAuto} reaches a median of 0.027~iter/s and 0.031
at eight times the budget.
\ours{} pays these evaluations once in offline training, while
\texttt{random} and \texttt{HetAuto} pay them again at every reconfiguration.

\PP{Reconfiguration overhead}
The measurement adds two A6000 GPUs twice to a node of two RTX~PRO~6000 GPUs
on the hardware cluster.
Each event runs live, with no checkpoint and no restart, and is repeated four
times.
Reconfiguration stops training for 19.2 to 21.5 sec end to end, and migrating
state is the largest step of the work.
The controller re-invokes the planner in 2.16 sec, the workers release their
resources in 0.4 to 0.8 sec, and the process group rebuild takes 2.6 to
3.1 sec.
The stage transfer of \autoref{sec:Reconfiguration} then hands 5.2\,GB of
model state per rank onto the added node in about 4.0 sec, never leaving GPU
memory for storage.
Building the engine on the new topology costs a further 2.1 to 2.7 sec.

%% file: 08_discussion.tex
\section{Limitations and Discussion}
\label{sec:discussion}

\PP{Diverse optimization objectives} Sailor~\cite{Sailor} pioneers support for various optimization objectives beyond maximum throughput. A
notable example is monetary cost per training iteration, which is essential
for cost-effective cloud deployments. Additionally, the arithmetic auto-fill mechanism should be modified to explore DP degrees in descending order. This allows \ours{} to account for the increased hourly burn rate of larger clusters and select the most cost-effective replica count that avoids the scaling tax of sub-linear performance gains~\cite{Sailor}. %
\ours{} then can treat performance-per-dollar as the primary benefit when deciding whether
to invest more search effort into a particular subspace.

\PP{Preemptible instances} Preemptible instances~\cite{aws_spot, google_spot, azure_spot} offer significant discounts
on GPU resources, but are subject to frequent revocation by high-priority
demands. To mitigate interruption impacts, cloud providers issue notices with
a brief grace period (e.g., 2 mins on AWS~\cite{aws_notice}, 30 secs on Google Cloud~\cite{gcp_notice}). We
believe \ours{} is well-suited for AWS in all of our evaluated models
and clusters, as worst-case planning times remain %
10 and 20 secs, respectively. Google Cloud presents a challenge, as memory state
migration for the largest evaluated model may exceed the 30-sec window. To
handle such cases, \ours{} could be augmented with on-demand CPU instances to
use fast, iteration-based checkpointing schemes like
Checkmate~\cite{checkmate}.

%% file: 09_related.tex
\section{Related Work}
\label{sec:related}

\PP{Parallelism planner} While we have extensively compared \ours{} with
AMP~\cite{amp}, Metis~\cite{Metis},
HexiScale~\cite{HexiScale}, Espresso~\cite{Espresso}, and
Sailor~\cite{Sailor} throughout this paper, there are a few other planners
including Galvatron~\cite{galvatron}, which applies dynamic programming over
a layered search space, and Cephalo~\cite{cephalo}, which targets
heterogeneous clusters using a profile-guided approach. All of these systems
solve each search problem in isolation. \ours{} instead employs RL to 
learn from prior optimization experiences and produce optimal plans 
near-instantaneously via model inference.

\PP{Reconfiguration and resilience} Some prior efforts
have established principled mechanisms for state transition during
parallelism reconfiguration. Varuna~\cite{varuna} enables elastic training by
dynamically scaling pipeline and data parallelism. ReCycle~\cite{recycle}
adaptively reshapes pipeline stages during resource shifts, while
Oobleck~\cite{oobleck} leverages pre-calculated pipeline templates for rapid
recovery from node failures. 
To accelerate state transition,
Tenplex~\cite{tenplex} and Universal
Checkpointing~\cite{universal_checkpointing} introduce specialized
abstractions that decouple model states from specific hardware topologies.
HotSPa~\cite{ge2024enabling} further enables runtime switching between hybrid
parallelism plans to mitigate load imbalances caused by varying sequence
lengths. None of these works, however, focuses on fast searching for an optimal
parallelism plan like \ours{}.

\PP{Learning-based device placement}
RL has been applied to the device
placement for computation graphs.
Mirhoseini et al.~\cite{hierarchicalRL} use a
hierarchical RL approach that groups operations
and assigns groups to devices.
Placeto~\cite{placeto} learns placement policies
that transfer across unseen model graphs.
HSDAG~\cite{HSDAG} introduces structure-aware
graph representations to improve placement
decisions.
FlexFlow~\cite{flexflow} searches per-operator strategies over four
dimensions and does express data parallelism, but it targets homogeneous
compositions and does not search device type.
These methods operate at the granularity of 
individual operations and fail to express 
data parallelism or generalize across 
cluster configurations.
\ours{} formulates planning at the template level,
where each action selects a PP degree, GPU
type, and TP degree, while 
deriving the remaining arithmetic decisions
deterministically. This allows \ours{} to capture
diverse parallelism dimensions in use and 
generalize across heterogeneous cluster configurations.

%% file: 10_conclusion.tex
\section{Conclusion}

\ours{} is a learning-based planner for
heterogeneous 4D parallelism in distributed model
training.
\ours{} decomposes the planning problem into
structural decisions handled by a learned policy
and arithmetic decisions that follow by rule or
from a small priced candidate set, reducing the
effective search space while preserving plan
quality.
A template-level autoregressive construction
policy with feasibility masking explores the full
space of valid configurations without heuristic
pruning.
A state of type fractions, device properties and
the cluster scale lets one trained policy plan for
a composition it has not seen.
Evaluation on clusters with up to 4 GPU types
shows that \ours{} achieves up to 84.5\% higher
throughput than state-of-the-art planners while
robust to dynamic GPU compositions. %

%% file: Paper.bbl
\begin{thebibliography}{10}

\bibitem{placeto}
Ravichandra Addanki, Shaileshh~Bojja Venkatakrishnan, Shreyan Gupta, Hongzi
  Mao, and Mohammad Alizadeh.
\newblock Placeto: learning generalizable device placement algorithms for
  distributed machine learning.
\newblock In {\em Advances in Neural Information Processing Systems},
  volume~32, Red Hook, NY, USA, 2019.

\bibitem{varuna}
Sanjith Athlur, Nitika Saran, Muthian Sivathanu, Ramachandran Ramjee, and Nipun
  Kwatra.
\newblock {Varuna: scalable, low-cost training of massive deep learning
  models}.
\newblock In {\em Proceedings of the Seventeenth European Conference on
  Computer Systems}, page 472–487. Association for Computing Machinery, 2022.

\bibitem{azure_spot}
Microsoft Azure.
\newblock Azure {Spot} {Virtual} {Machines}.
\newblock
  \url{https://azure.microsoft.com/en-us/products/virtual-machines/spot/},
  2026.

\bibitem{checkmate}
Ankit Bhardwaj, Weiyang Wang, Jeremy Carin, Adam Belay, and Manya Ghobadi.
\newblock Checkmate: zero performance over head model checkpointing via network
  gradient replication.
\newblock In {\em 23rd USENIX Symposium on Networked Systems Design and
  Implementation (NSDI 26)}, USA, 2026. USENIX Association.

\bibitem{gandiva-fair}
Shubham Chaudhary, Ramachandran Ramjee, Muthian Sivathanu, Nipun Kwatra, and
  Srinidhi Viswanatha.
\newblock Balancing efficiency and fairness in heterogeneous gpu clusters for
  deep learning.
\newblock In {\em Proceedings of the Fifteenth European Conference on Computer
  Systems}. Association for Computing Machinery, 2020.

\bibitem{gcp_notice}
Google Cloud.
\newblock Google cloud, preemption process on {Compute} {Engine}.
\newblock
  \url{https://cloud.google.com/compute/docs/instances/preemptible#preemption-process},
  2026.

\bibitem{googlecloud_gpus}
Google Cloud.
\newblock Gpus on compute engine.
\newblock \url{https://cloud.google.com/compute/docs/gpus}, 2026.

\bibitem{google_spot}
Google Cloud.
\newblock Spot {VMs} on {Google} {Cloud}.
\newblock \url{https://cloud.google.com/spot-vms}, 2026.

\bibitem{deepseekv3}
DeepSeek-AI, Aixin Liu, Bei Feng, Bing Xue, Bingxuan Wang, Bochao Wu, Chengda
  Lu, Chenggang Zhao, Chengqi Deng, Chenyu Zhang, Chong Ruan, Damai Dai, Daya
  Guo, Dejian Yang, Deli Chen, Dongjie Ji, Erhang Li, Fangyun Lin, Fucong Dai,
  Fuli Luo, and et~al.
\newblock Deepseek-v3 technical report.
\newblock {\em arXiv preprint arXiv:2412.19437}, 2024.

\bibitem{parcae}
Jiangfei Duan, Ziang Song, Xupeng Miao, Xiaoli Xi, Dahua Lin, Harry Xu, Minjia
  Zhang, and Zhihao Jia.
\newblock Parcae: Proactive, {Liveput-Optimized} {DNN} training on preemptible
  instances.
\newblock In {\em 21st USENIX Symposium on Networked Systems Design and
  Implementation (NSDI 24)}, pages 1121--1139, Santa Clara, CA, April 2024.
  USENIX Association.

\bibitem{HSDAG}
Shukai Duan, Heng Ping, Nikos Kanakaris, Xiongye Xiao, Panagiotis Kyriakis,
  Nesreen~K. Ahmed, Peiyu Zhang, Guixiang Ma, Mihai Capot\u{a}, Shahin
  Nazarian, Theodore~L. Willke, and Paul Bogdan.
\newblock A structure-aware framework for learning device placements on
  computation graphs.
\newblock In A.~Globerson, L.~Mackey, D.~Belgrave, A.~Fan, U.~Paquet,
  J.~Tomczak, and C.~Zhang, editors, {\em Advances in Neural Information
  Processing Systems}, volume~37, pages 81748--81772, 2024.

\bibitem{switchformer}
William Fedus, Barret Zoph, and Noam Shazeer.
\newblock Switch transformers: Scaling to trillion parameter models with simple
  and efficient sparsity.
\newblock {\em Journal of Machine Learning Research}, 23(120):1--39, 2022.

\bibitem{recycle}
Swapnil Gandhi, Mark Zhao, Athinagoras Skiadopoulos, and Christos Kozyrakis.
\newblock Recycle: Resilient training of large dnns using pipeline adaptation.
\newblock In {\em Proceedings of the ACM SIGOPS 30th Symposium on Operating
  Systems Principles}, SOSP '24, page 211–228, New York, NY, USA, 2024.
  Association for Computing Machinery.

\bibitem{ge2024enabling}
Hao Ge, Fangcheng Fu, Haoyang Li, Xuanyu Wang, Sheng Lin, Yujie Wang, Xiaonan
  Nie, Hailin Zhang, Xupeng Miao, and Bin Cui.
\newblock Enabling parallelism hot switching for efficient training of large
  language models.
\newblock In {\em Proceedings of the ACM SIGOPS 30th Symposium on Operating
  Systems Principles}, SOSP '24, page 178–194, New York, NY, USA, 2024.
  Association for Computing Machinery.

\bibitem{llama3}
Aaron Grattafiori, Abhimanyu Dubey, Abhinav Jauhri, Abhinav Pandey, Abhishek
  Kadian, Ahmad Al-Dahle, Aiesha Letman, Akhil Mathur, Alan Schelten, Alex
  Vaughan, et~al.
\newblock The llama 3 herd of models.
\newblock {\em arXiv preprint arXiv:2407.21783}, 2024.

\bibitem{cephalo}
Runsheng~Benson Guo, Utkarsh Anand, Arthur Chen, and Khuzaima Daudjee.
\newblock Cephalo: Harnessing heterogeneous gpu clusters for training
  transformer models.
\newblock In {\em Proceedings of the 39th ACM International Conference on
  Supercomputing}, ICS '25, page 368–383, New York, NY, USA, 2025.
  Association for Computing Machinery.

\bibitem{hu2025demystifying}
Zhiyi Hu, Siyuan Shen, Tommaso Bonato, Sylvain Jeaugey, Cedell Alexander, Eric
  Spada, James Dinan, Jeff Hammond, and Torsten Hoefler.
\newblock {Demystifying NCCL: An in-depth analysis of GPU communication
  protocols and algorithms}.
\newblock In {\em 2025 IEEE Symposium on High-Performance Interconnects}, IEEE,
  pages 48--59. IEEE, 2025.

\bibitem{gpipe}
Yanping Huang, Youlong Cheng, Ankur Bapna, Orhan Firat, Dehao Chen, Mia Chen,
  HyoukJoong Lee, Jiquan Ngiam, Quoc~V Le, Yonghui Wu, and zhifeng Chen.
\newblock {GPipe: Efficient Training of Giant Neural Networks using Pipeline
  Parallelism}.
\newblock In H.~Wallach, H.~Larochelle, A.~Beygelzimer, F.~d\textquotesingle
  Alch\'{e}-Buc, E.~Fox, and R.~Garnett, editors, {\em Advances in Neural
  Information Processing Systems}, volume~32, 2019.

\bibitem{GPT-Neo}
HuggingFace.
\newblock gpt-neo.
\newblock \url{https://huggingface.co/docs/transformers/en/model_doc/gpt_neo},
  2021.

\bibitem{tutel}
Changho Hwang, Wei Cui, Yifan Xiong, Ziyue Yang, Ze~Liu, Han Hu, Zilong Wang,
  Rafael Salas, Jithin Jose, Prabhat Ram, Joe Chau, Peng Cheng, Fan Yang, Mao
  Yang, and Yongqiang Xiong.
\newblock {Tutel}: Adaptive mixture-of-experts at scale.
\newblock In {\em Proceedings of Machine Learning and Systems}, volume~5, pages
  269--287, 2023.

\bibitem{oobleck}
Insu Jang, Zhenning Yang, Zhen Zhang, Xin Jin, and Mosharaf Chowdhury.
\newblock {Oobleck: Resilient Distributed Training of Large Models Using
  Pipeline Templates}.
\newblock In {\em Proceedings of the 29th Symposium on Operating Systems
  Principles}, SOSP '23, page 382–395, New York, NY, USA, 2023. Association
  for Computing Machinery.

\bibitem{philly}
Myeongjae Jeon, Shivaram Venkataraman, Amar Phanishayee, Junjie Qian, Wencong
  Xiao, and Fan Yang.
\newblock Analysis of {Large-Scale} {Multi-Tenant} {GPU} clusters for {DNN}
  training workloads.
\newblock In {\em 2019 USENIX Annual Technical Conference (USENIX ATC 19)},
  pages 947--960. USENIX Association, 2019.

\bibitem{whale}
Xianyan Jia, Le~Jiang, Ang Wang, Wencong Xiao, Ziji Shi, Jie Zhang, Xinyuan Li,
  Langshi Chen, Yong Li, Zhen Zheng, Xiaoyong Liu, and Wei Lin.
\newblock {Whale: Efficient Giant Model Training over Heterogeneous GPUs}.
\newblock In {\em 2022 USENIX Annual Technical Conference (USENIX ATC 22)},
  pages 673--688. USENIX Association, 2022.

\bibitem{flexflow}
Zhihao Jia, Matei Zaharia, and Alex Aiken.
\newblock Beyond data and model parallelism for deep neural networks.
\newblock In A.~Talwalkar, V.~Smith, and M.~Zaharia, editors, {\em Proceedings
  of Machine Learning and Systems}, volume~1, pages 1--13, 2019.

\bibitem{mixtral}
Albert~Q. Jiang, Alexandre Sablayrolles, Antoine Roux, Arthur Mensch, Blanche
  Savary, Chris Bamford, Devendra~Singh Chaplot, Diego de~las Casas, Emma~Bou
  Hanna, Florian Bressand, Gianna Lengyel, Guillaume Bour, Guillaume Lample,
  Lélio~Renard Lavaud, Lucile Saulnier, Marie-Anne Lachaux, Pierre Stock,
  Sandeep Subramanian, Sophia Yang, Szymon Antoniak, Teven~Le Scao, Théophile
  Gervet, Thibaut Lavril, Thomas Wang, Timothée Lacroix, and William~El Sayed.
\newblock Mixtral of experts.
\newblock {\em arXiv preprint arXiv:2401.04088}, 2024.

\bibitem{megascale}
Ziheng Jiang, Haibin Lin, Yinmin Zhong, Qi~Huang, Yangrui Chen, Zhi Zhang,
  Yanghua Peng, Xiang Li, Cong Xie, Shibiao Nong, Yulu Jia, Sun He, Hongmin
  Chen, Zhihao Bai, Qi~Hou, Shipeng Yan, Ding Zhou, Yiyao Sheng, Zhuo Jiang,
  Haohan Xu, Haoran Wei, Zhang Zhang, Pengfei Nie, Leqi Zou, Sida Zhao, Liang
  Xiang, Zherui Liu, Zhe Li, Xiaoying Jia, Jianxi Ye, Xin Jin, and Xin Liu.
\newblock {MegaScale: Scaling Large Language Model Training to More Than 10,000
  GPUs}.
\newblock In {\em 21st USENIX Symposium on Networked Systems Design and
  Implementation (NSDI 24)}, pages 745--760, Santa Clara, CA, April 2024.
  USENIX Association.

\bibitem{megascalemoe}
Chao Jin, Ziheng Jiang, Zhihao Bai, Zheng Zhong, Juncai Liu, Xiang Li, Ningxin
  Zheng, Xi~Wang, Cong Xie, Qi~Huang, Wen Heng, Yiyuan Ma, Wenlei Bao, Size
  Zheng, Xuegui Zheng, Yanghua Peng, Haibin Lin, Xuanzhe Liu, Xin Jin, and Xin
  Liu.
\newblock Megascale-moe: Large-scale communication-efficient training of
  mixture-of-experts models in production.
\newblock In {\em Proceedings of the 21st European Conference on Computer
  Systems}, page 366–382. Association for Computing Machinery, 2026.

\bibitem{gshard}
Dmitry Lepikhin, HyoukJoong Chung, Ian Fedorov, Younes Souli, Sabira Shukurova,
  Cheng-An Kan, Ye~Zhu, Quoc~V. Le, Yonghui Wu, and Zhifeng Chen.
\newblock {GShard: Scaling Giant Models with Conditional Computation and
  Automatic Sharding}.
\newblock In {\em The 9th International Conference on Learning
  Representations}, 2021.

\bibitem{amp}
Dacheng Li, Hongyi Wang, Eric Xing, and Hao Zhang.
\newblock Amp: Automatically finding model parallel strategies with
  heterogeneity awareness.
\newblock In S.~Koyejo, S.~Mohamed, A.~Agarwal, D.~Belgrave, K.~Cho, and A.~Oh,
  editors, {\em Advances in Neural Information Processing Systems}, volume~35,
  pages 6630--6639, 2022.

\bibitem{ASI}
Suyi Li, Lingyun Yang, Haoxuan Yu, Sheng Yao, Tianyuan Wu, Xiaoxiao Jiang,
  Hanfeng Lu, Kangjin Wang, Chenhao Wang, Shenglin Xu, Lun Wang, Qingyang Duan,
  Shenghao Liang, Xiu Lin, Meng Zhang, Wenchao Wu, Yinghao Yu, Guodong Yang,
  Liping Zhang, and Wei Wang.
\newblock Heterogeneity at hyperscale: Characterization and scheduling of large
  production {AI} clusters at {Alibaba}.
\newblock In {\em 20th USENIX Symposium on Operating Systems Design and
  Implementation (OSDI 26)}, pages 2187--2203, 2026.

\bibitem{gpunion2025}
Yufang Li, Yuanbo Zhang, Hanlong Liao, Deke Guo, and Guoming Tang.
\newblock Gpunion: Autonomous gpu sharing on campus.
\newblock In {\em Proceedings of the 24th ACM Workshop on Hot Topics in
  Networks}, HotNets '25, page 96–103. Association for Computing Machinery,
  2025.

\bibitem{universal_checkpointing}
Xinyu Lian, Sam~Ade Jacobs, Lev Kurilenko, Masahiro Tanaka, Stas Bekman,
  Olatunji Ruwase, and Minjia Zhang.
\newblock Universal checkpointing: a flexible and efficient distributed
  checkpointing system for large-scale dnn training with reconfigurable
  parallelism.
\newblock In {\em 2025 USENIX Annual Technical Conference (USENIX ATC 25)},
  pages 1519--1534. USENIX Association, 2025.

\bibitem{moefolding}
Dennis Liu, Zijie Yan, Xin Yao, Tong Liu, Vijay Korthikanti, Evan Wu, Shiqing
  Fan, Gao Deng, Hongxiao Bai, Jianbin Chang, Ashwath Aithal, Michael Andersch,
  Mohammad Shoeybi, Jiajie Yao, Chandler Zhou, David Wu, Xipeng Li, and June
  Yang.
\newblock {MoE} parallel folding: Heterogeneous parallelism mappings for
  efficient large-scale {MoE} model training with megatron core.
\newblock {\em arXiv preprint arXiv:2504.14960}, 2025.

\bibitem{hexamoe}
Shuqing Luo, Jie Peng, Pingzhi Li, Hanrui Wang, and Tianlong Chen.
\newblock {Hexa-MoE}: Efficient and heterogeneous-aware training for
  mixture-of-experts.
\newblock {\em arXiv preprint arXiv:2411.01288}, 2024.

\bibitem{galvatron}
Xupeng Miao, Yujie Wang, Youhe Jiang, Chunan Shi, Xiaonan Nie, Hailin Zhang,
  and Bin Cui.
\newblock Galvatron: Efficient transformer training over multiple gpus using
  automatic parallelism.
\newblock {\em Proc. VLDB Endow.}, 16(3):470–479, November 2022.

\bibitem{hierarchicalRL}
Azalia Mirhoseini, Anna Goldie, Hieu Pham, Benoit Steiner, Quoc~V Le, and Jeff
  Dean.
\newblock A hierarchical model for device placement.
\newblock In {\em The 6th International Conference on Learning
  Representations}, 2018.

\bibitem{pipedream}
Deepak Narayanan, Aaron Harlap, Amar Phanishayee, Vivek Seshadri, Nikhil~R.
  Devanur, Gregory~R. Ganger, Phillip~B. Gibbons, and Matei Zaharia.
\newblock Pipedream: generalized pipeline parallelism for dnn training.
\newblock In {\em Proceedings of the 27th ACM Symposium on Operating Systems
  Principles}, SOSP '19, page 1–15, New York, NY, USA, 2019. Association for
  Computing Machinery.

\bibitem{gavel}
Deepak Narayanan, Keshav Santhanam, Fiodar Kazhamiaka, Amar Phanishayee, and
  Matei Zaharia.
\newblock {Heterogeneity-Aware Cluster Scheduling Policies for Deep Learning
  Workloads}.
\newblock In {\em 14th USENIX Symposium on Operating Systems Design and
  Implementation (OSDI 20)}, pages 481--498, 2020.

\bibitem{Oh2018SIL}
Junhyuk Oh, Yijie Guo, Satinder Singh, and Honglak Lee.
\newblock Self-imitation learning.
\newblock In Jennifer Dy and Andreas Krause, editors, {\em Proceedings of the
  35th International Conference on Machine Learning}, volume~80, pages
  3878--3887. PMLR, 2018.

\bibitem{HetAuto}
Guicheng Qi, Junwei Su, Liqi Yang, Tao Li, Tingwen Xie, Yerui Sun, Yuchen Xie,
  and Chuan Wu.
\newblock {HetAuto: Cross-Cluster Auto-Parallelism for Heterogeneous
  Distributed Training}.
\newblock In {\em Proceedings of the 21st European Conference on Computer
  Systems}, page 759–779, Edinburgh, Scotland, UK, 2026. ACM.

\bibitem{deepspeedmoe}
Samyam Rajbhandari, Conglong Li, Zhewei Yao, Minjia Zhang, Reza~Yazdani
  Aminabadi, Ammar~Ahmad Awan, Jeff Rasley, and Yuxiong He.
\newblock {D}eep{S}peed-{M}o{E}: Advancing mixture-of-experts inference and
  training to power next-generation {AI} scale.
\newblock In {\em Proceedings of the 39th International Conference on Machine
  Learning}, volume 162, pages 18332--18346. PMLR, 2022.

\bibitem{deepspeed}
Jeff Rasley, Samyam Rajbhandari, Olatunji Ruwase, and Yuxiong He.
\newblock {DeepSpeed: System Optimizations Enable Training Deep Learning Models
  with Over 100 Billion Parameters}.
\newblock In {\em Proceedings of the 26th ACM SIGKDD International Conference
  on Knowledge Discovery \& Data Mining}, KDD '20, page 3505–3506, New York,
  NY, USA, 2020. Association for Computing Machinery.

\bibitem{switchML}
Amedeo Sapio, Marco Canini, Chen-Yu Ho, Jacob Nelson, Panos Kalnis, Changhoon
  Kim, Arvind Krishnamurthy, Masoud Moshref, Dan Ports, and Peter Richtarik.
\newblock {Scaling Distributed Machine Learning with {In-Network} Aggregation}.
\newblock In {\em 18th USENIX Symposium on Networked Systems Design and
  Implementation (NSDI 21)}, pages 785--808. USENIX Association, April 2021.

\bibitem{PPO}
John Schulman, Filip Wolski, Prafulla Dhariwal, Alec Radford, and Oleg Klimov.
\newblock Proximal policy optimization algorithms.
\newblock {\em arXiv preprint arXiv:1707.06347}, 2017.

\bibitem{aws_spot}
Amazon~Web Services.
\newblock Amazon {EC2} {Spot} {Instances}.
\newblock \url{https://aws.amazon.com/ec2/spot/}, 2026.

\bibitem{aws_notice}
Amazon~Web Services.
\newblock Amazon web services, spot instance interruption notices.
\newblock
  \url{https://docs.aws.amazon.com/AWSEC2/latest/UserGuide/spot-interruptions.html},
  2026.

\bibitem{grpo}
Zhihong Shao, Peiyi Wang, Qihao Zhu, Runxin Xu, Junxiao Song, Xiao Bi, Haowei
  Zhang, Mingchuan Zhang, YK~Li, Yang Wu, et~al.
\newblock Deepseekmath: Pushing the limits of mathematical reasoning in open
  language models.
\newblock {\em arXiv preprint arXiv:2402.03300}, 2024.

\bibitem{megatron-lm}
Mohammad Shoeybi, Mostofa Patwary, Raul Puri, Patrick LeGresley, Jared Casper,
  and Bryan Catanzaro.
\newblock Megatron-lm: Training multi-billion parameter language models using
  model parallelism.
\newblock {\em arXiv preprint arXiv:1909.08053}, 2020.

\bibitem{Sailor}
Foteini Strati, Zhendong Zhang, George Manos, Ixeia~S\'{a}nchez P\'{e}riz,
  Qinghao Hu, Tiancheng Chen, Berk Buzcu, Song Han, Pamela Delgado, and Ana
  Klimovic.
\newblock Sailor: Automating distributed training over dynamic, heterogeneous,
  and geo-distributed clusters.
\newblock In {\em Proceedings of the ACM SIGOPS 31st Symposium on Operating
  Systems Principles}, SOSP '25, page 204–220, New York, NY, USA, 2025.
  Association for Computing Machinery.

\bibitem{bamboo}
John Thorpe, Pengzhan Zhao, Jonathan Eyolfson, Yifan Qiao, Zhihao Jia, Minjia
  Zhang, Ravi Netravali, and Guoqing~Harry Xu.
\newblock {Bamboo: Making Preemptible Instances Resilient for Affordable
  Training of Large DNNs}.
\newblock In {\em 20th USENIX Symposium on Networked Systems Design and
  Implementation (NSDI 23)}, pages 497--513. USENIX Association, April 2023.

\bibitem{Metis}
Taegeon Um, Byungsoo Oh, Minyoung Kang, Woo-Yeon Lee, Goeun Kim, Dongseob Kim,
  Youngtaek Kim, Mohd Muzzammil, and Myeongjae Jeon.
\newblock {Metis: Fast Automatic Distributed Training on Heterogeneous {GPUs}}.
\newblock In {\em 2024 USENIX Annual Technical Conference (USENIX ATC 24)},
  pages 563--578. USENIX Association, 2024.

\bibitem{tenplex}
Marcel Wagenl\"{a}nder, Guo Li, Bo~Zhao, Luo Mai, and Peter Pietzuch.
\newblock {Tenplex: Dynamic Parallelism for Deep Learning using Parallelizable
  Tensor Collections}.
\newblock In {\em Proceedings of the ACM SIGOPS 30th Symposium on Operating
  Systems Principles}, SOSP '24, page 195–210, New York, NY, USA, 2024.
  Association for Computing Machinery.

\bibitem{gemini-checkpoint}
Zhuang Wang, Zhen Jia, Shuai Zheng, Zhen Zhang, Xinwei Fu, T.~S.~Eugene Ng, and
  Yida Wang.
\newblock Gemini: Fast failure recovery in distributed training with in-memory
  checkpoints.
\newblock In {\em Proceedings of the 29th Symposium on Operating Systems
  Principles}, SOSP '23, page 364–381, New York, NY, USA, 2023. Association
  for Computing Machinery.

\bibitem{MLaaS}
Qizhen Weng, Wencong Xiao, Yinghao Yu, Wei Wang, Cheng Wang, Jian He, Yong Li,
  Liping Zhang, Wei Lin, and Yu~Ding.
\newblock {MLaaS} in the wild: Workload analysis and scheduling in
  {Large-Scale} heterogeneous {GPU} clusters.
\newblock In {\em 19th USENIX Symposium on Networked Systems Design and
  Implementation (NSDI 22)}, pages 945--960, Renton, WA, April 2022. USENIX
  Association.

\bibitem{hetermoe}
Yongji Wu, Xueshen Liu, Shuowei Jin, Ceyu Xu, Feng Qian, Z.~Morley Mao, Matthew
  Lentz, Danyang Zhuo, and Ion Stoica.
\newblock {HeterMoE}: Efficient training of mixture-of-experts models on
  heterogeneous gpus.
\newblock {\em arXiv preprint arXiv:2504.03871}, 2025.

\bibitem{HexiScale}
Ran Yan, YOUHE JIANG, Xiaonan Nie, Fangcheng Fu, Bin CUI, and Binhang Yuan.
\newblock Hexiscale: Facilitating large language model training over
  heterogeneous hardware.
\newblock In A.~Chowdhery and Z.~Jia, editors, {\em Proceedings of Machine
  Learning and Systems}, volume~8, pages 821--842. MLSys, 2026.

\bibitem{Qwen}
An~Yang, Anfeng Li, Baosong Yang, Beichen Zhang, Binyuan Hui, Bo~Zheng, Bowen
  Yu, Chang Gao, Chengen Huang, Chenxu Lv, Chujie Zheng, Dayiheng Liu, Fan
  Zhou, Fei Huang, Feng Hu, Hao Ge, Haoran Wei, Huan Lin, Jialong Tang, Jian
  Yang, Jianhong Tu, Jianwei Zhang, Jianxin Yang, Jiaxi Yang, Jing Zhou,
  Jingren Zhou, Junyang Lin, Kai Dang, Keqin Bao, Kexin Yang, Le~Yu, Lianghao
  Deng, Mei Li, Mingfeng Xue, Mingze Li, Pei Zhang, Peng Wang, Qin Zhu, Rui
  Men, Ruize Gao, Shixuan Liu, Shuang Luo, Tianhao Li, Tianyi Tang, Wenbiao
  Yin, Xingzhang Ren, Xinyu Wang, Xinyu Zhang, Xuancheng Ren, Yang Fan, Yang
  Su, Yichang Zhang, Yinger Zhang, Yu~Wan, Yuqiong Liu, Zekun Wang, Zeyu Cui,
  Zhenru Zhang, Zhipeng Zhou, and Zihan Qiu.
\newblock Qwen3 technical report.
\newblock {\em arXiv preprint arXiv:2505.09388}, 2025.

\bibitem{pcgrad}
Tianhe Yu, Saurabh Kumar, Abhishek Gupta, Sergey Levine, Karol Hausman, and
  Chelsea Finn.
\newblock Gradient surgery for multi-task learning.
\newblock In H.~Larochelle, M.~Ranzato, R.~Hadsell, M.F. Balcan, and H.~Lin,
  editors, {\em Advances in Neural Information Processing Systems}, volume~33,
  pages 5824--5836, 2020.

\bibitem{alpa}
Lianmin Zheng, Zhuohan Li, Hao Zhang, Yonghao Zhuang, Zhifeng Chen, Yanping
  Huang, Yida Wang, Yuanzhong Xu, Danyang Zhuo, Eric~P. Xing, Joseph~E.
  Gonzalez, and Ion Stoica.
\newblock Alpa: Automating inter- and {Intra-Operator} parallelism for
  distributed deep learning.
\newblock In {\em 16th USENIX Symposium on Operating Systems Design and
  Implementation (OSDI 22)}, pages 559--578, 2022.

\bibitem{Espresso}
Qiannan Zhou, Fei Xu, Lingxuan Weng, Ruixing Li, Xudong Wu, Li~Chen, Zhi Zhou,
  and Fangming Liu.
\newblock Espresso: Cost-efficient large model training by exploiting gpu
  heterogeneity in the cloud.
\newblock In {\em IEEE INFOCOM 2025 - IEEE Conference on Computer
  Communications}, IEEE, pages 1--10, 2025.

\end{thebibliography}
